\documentclass{article}
\usepackage[margin=1in]{geometry}
\usepackage[hidelinks]{hyperref}
\usepackage{amsmath}
\usepackage{amssymb}
\usepackage{parskip} 
\usepackage{natbib}
\usepackage{mathrsfs}
\usepackage{pdfpages}
\usepackage{bm}
\usepackage{booktabs}
\usepackage{graphics}
\usepackage{multicol}
\usepackage[bf]{caption}
\usepackage{multirow}
\usepackage{subcaption}
\usepackage{floatrow}
\usepackage{comment}
\usepackage{amsthm}
\usepackage[titletoc]{appendix}
\usepackage{setspace}
\usepackage{mwe}  
\usepackage{color}

\usepackage{authblk}

\title{
Multi-porous extension of anisotropic poroelasticity: consolidation and related coefficients
}

\author[1, 2]{\normalsize Filip P. Adamus}
\author[1]{\normalsize David Healy}
\author[2]{\normalsize Philip G. Meredith}
\author[2]{\normalsize Thomas M. Mitchell}
\author[2]{\normalsize Ashley Stanton-Yonge}
\affil[1]{\footnotesize {School of Geosciences, University of Aberdeen, Aberdeen, UK}}
\affil[2]{\footnotesize {Department of Earth Sciences, University College London, London, UK \\ \linebreak
\[adamusfp@gmail.com \quad d.healy@abdn.ac.uk \quad p.meredith@ucl.ac.uk \quad tom.mitchell@ucl.ac.uk \quad ashley.sesnic.18@ucl.ac.uk\]}}

\date{}

\begin{document}
\maketitle

\section*{\centering \small{Abstract}}
\small{
We propose the generalisation of the anisotropic poroelasticity theory. At a large scale, a medium is viewed as quasi-static, which is the original assumption of Biot. At a smaller scale, we distinguish different porosity clusters (sets of pores or fractures) that are characterized by various fluid pressures, which is the original poroelastic extension of Aifantis. In consequence, both instantaneous and time-dependent deformation lead to fluid content variations that are different in each cluster. We present the equations for such phenomena, where the anisotropic properties of both the solid matrix and pore sets are assumed. Novel poroelastic coefficients that relate solid and fluid phases in our extension are proposed, and their physical meaning is determined. To demonstrate the utility of our equations and emphasize the meaning of new coefficients, we perform numerical simulations of a triple-porosity consolidation. These simulations reveal positive pore pressure transients in the drained behaviour of weakly connected pore sets, and these may result in mechanical weakening of the material. 
}
\\ \\ 
{\bf{Keywords:}} Anisotropy, Fractures, Multiple-permeability, Multiple-porosity, Poroelasticity, Rock mechanics.

\newpage
\section{Introduction}
The deformation of a porous medium containing fluids can be described using equations proposed by Maurice Anthony Biot; an applied physicist born in Belgium. His phenomenological theory relates strains of a solid phase with displacements of fluids. A medium is viewed at a macroscopic, bulk scale, where all pores are treated as interconnected. Constant fluid pressure throughout the medium and a unique measure of the fluid content change is adopted. His theory of three-dimensional consolidation describes time-dependent deformation. It assumes quasi-static stress conditions, incompressible fluid, and flow obeying Darcy's law. In 1935, the isotropic version of consolidation was formulated by Biot in French. Six years later, the more rigorous treatment was written in English~\citep{Biot41}. Finally, Biot discussed a more general process of anisotropic consolidation~\citep{Biot55}. The developments from the aforementioned papers formulate the core of the so-called (quasi-static) poroelasticity. The quasi-static approach will also be considered in this paper.

Since the publication of the original theory of poroelastic consolidation, some generalisations were proposed. 
\citet{Biot56Visco} considered a viscoelastic extension. 
Further, \citet{Biot72} formulated the finite elastic description of porous structures.
Other researchers introduced more modifications. 
\citet{BookerSavvidou84} generalized the consolidation equations by including a thermal stress term.
Chemical effects were considered by~\citet{Sherwood93}.
Various impacts were combined to provide multi-physical formulations for porous materials by~\citet{TaronEtAl09}. They linked the usual mechanical and hydraulic poroelastic coupling with thermal and chemical effects. Note that all the above extensions go beyond the theory of poroelasticity. These generalisations constitute foundations of the theories of \textit{finite poroelasticity}, \textit{porothermoelasticity}, \textit{porochemoelasticity}, or \textit{porothermochemoelasticity}, respectively~\citep[Chapter 12,][]{Cheng16}. In this paper, however, we do not go beyond poroelasticity. Instead, we consider an extension within the frame of elastic, mechanical-hydraulic coupling.

Such an extension within Biot's poroelastic frame was first formulated by~\citet{Aifantis77,Aifantis80c,Aifantis80b,Aifantis80a}, a physicist born in Greece. His generalisation of consolidation equations was based on the idea of the so-called dual or multiple-porosity. A ''multi-porous'' medium consists of two or more sets of pores that are characterized by different fluid pressures and diffusions. In other words, the assumption of uniform pore pressure---originally introduced by Biot---is removed. It is important to mention that the concept of spatially varying pressure has been known long before the aforementioned work. \citet{BarenblattEtAl60} were the first to introduce the dual-porosity approach and \citet{WarrenRoot63} were the first to discuss its use in practice. However, these authors considered the problem of fluid diffusion only. As noticed by~\citet{ElsworthBai92}, until the work of~\citet{Aifantis77}, the hydrologic description was not coupled with mechanics since the stresses were assumed to be constant~\citep{Odeh65,DuguidLee77}. 
In the classical dual-porosity analysis of fluid diffusion, the fractures and solid matrix contribute to the overall behaviour in distinctly different ways due to decidedly different characteristics; the fractures commonly have high permeability and low storativity, whereas the solid matrix tends to have the opposite characteristics. Therefore, it is reasonable to expect the poromechanical contributions of the fractures to vary from that of the solid matrix.
The works by Aifantis led to the multi-porous extension of the isotropic poroelasticity theory. The aim of this paper is to generalise the formulations of Aifantis by considering the effect of anisotropy caused by the structure of the solid matrix and different sets of pores. 

Since the pioneering work of~\citet{Aifantis77}, the multi-porous (including dual-porosity) consolidation was discussed continuously by numerous researchers for over forty years (see revision of~\citet{AshworthDoster19}). First, \citet{WilsonAifantis82} determined the extended poroelastic coefficients in terms of volume fractions and various bulk moduli. Then, \citet{KhaledEtAl84} formulated the consolidation equations using coefficients of fluid content change utilized by Biot. \citet{ChoEtAl91} introduced anisotropy to the solid skeleton and one type of porosity, however, the poroelastic coefficients still remained isotropic. \citet{BerrymanWang95} proposed a novel representation of isotropic strain-stress relations, linked with symmetric coefficients $a_{ij}$.
They discussed the physical meaning of each coefficient. The alternative determination using the so-called uniform expansion thought experiment was discussed by~\citet{BerrymanPride02} and \citet{Berryman02}. Further,~\citet{Nguyen10} and~\citet{Mehrabian14} solved isotropic Mandel-type consolidation problems. The latter work provided explicit formulations for multi-porous consolidations. \citet{Mehrabian18} discussed isotropic strain-stress, stress-strain, and mixture relations in view of the multi-porous coefficients. Recently,~\citet{ZhangBorja21}, and~\citet{ZhangEtAl21} provided an anisotropic extension of the dual-porosity diffusion equation using stress-strain relations and various non-phenomenological methods that did not include Biot's fluid increment coefficient, $\zeta$. In this paper, our novel anisotropic extension is based on Biot's approach, where the fluid continuity equations~\citep{Biot41} are used. We present new anisotropic equations of multi-porous three-dimensional consolidation. Also, our anisotropic poroelastic coefficients differ from the ones of~\citet{ZhangBorja21} since they are derived from strain-stress relations without the assumption of superposition. Part of our physical determination methodology is similar to that of~\citet{BerrymanWang95}. 

Before we move to the theoretical derivations (Sections~\ref{sec3}-\ref{sec6}) followed by numerical examples (Section~\ref{sec:num}), we first sketch in the next section the idea of our anisotropic multi-porous extension and indicate its potential usage. Also, we refer to previous works that considered the practical aspects of dual porosity. 
\section{Potential usage of anisotropic multi-porous extension}\label{sec2}
Many porous materials, such as rocks, composites, or human tissues, exhibit anisotropic behaviour. The directional-dependent mechanical response of a medium can have various causes that are often combined. The anisotropy may be intrinsic (e.g., arrangement of crystals), induced by inhomogeneities (e.g., pores), or induced by stress (e.g., fractures that record deformation due to pre-existing loads). As mentioned in the previous section, we allow the two former origins of anisotropy in porous media. The intrinsic anisotropy is described by a solid matrix. If pores are not spherical and are not randomly oriented, then induced anisotropy is additionally present. 

\begin{figure}[!htbp]
\centering
\includegraphics[scale=0.5]{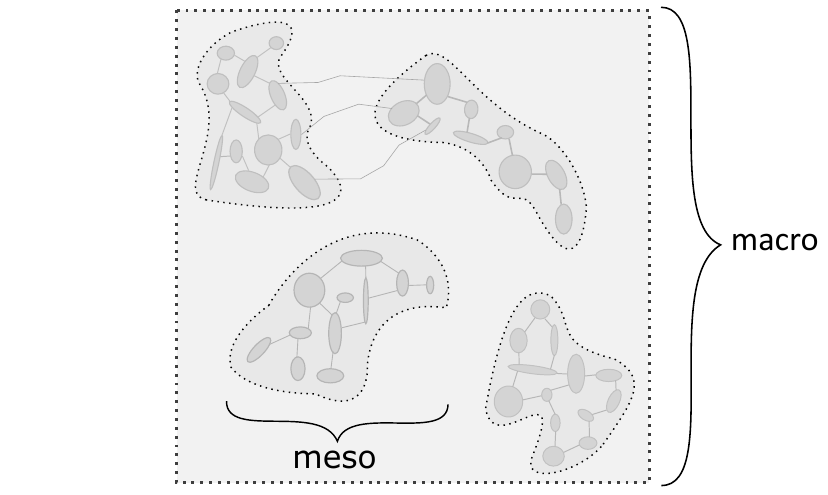}
\caption{\footnotesize{Schematic view of the multi-porous extension that considers a quasi-static medium at a macroscopic scale and distinguishes pore sets at a mesoscopic scale. Sets are allowed to be weakly connected (see upper part).}}
\label{fig:one}
\end{figure}

\begin{figure}[!htbp]
\centering
\begin{subfigure}{.35\textwidth}
\includegraphics[scale=0.35]{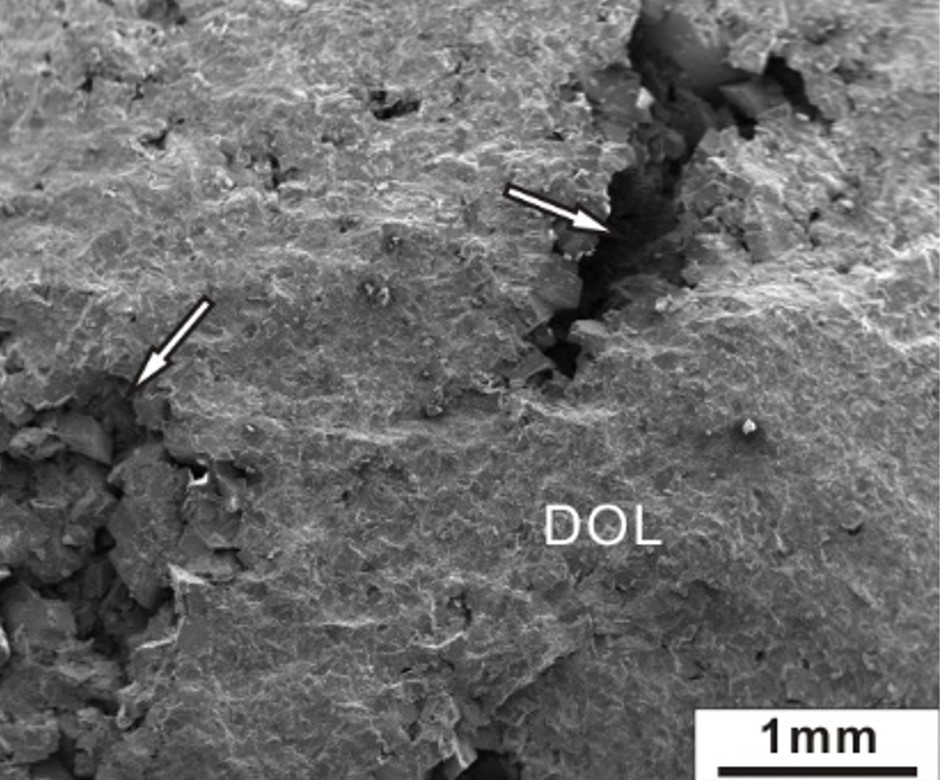}
\caption{\footnotesize{Hierarchical porosity: dolostone}}
\label{fig:a}
\end{subfigure}
 \qquad\qquad\vspace{0.5cm}
\begin{subfigure}{.34\textwidth}
\includegraphics[scale=0.323]{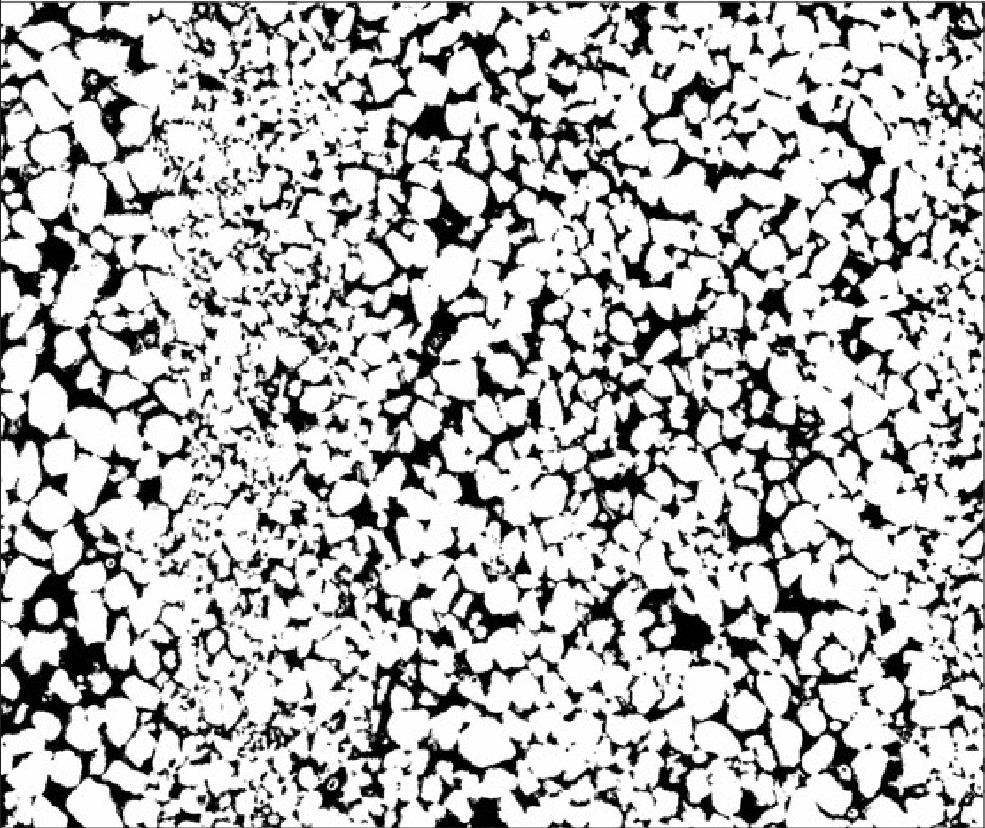}
\caption{\footnotesize{Hierarchical porosity: sandstone}}
\label{fig:b}
\end{subfigure}
\vspace{1cm}
\begin{subfigure}{.35\textwidth}
\includegraphics[scale=0.375]{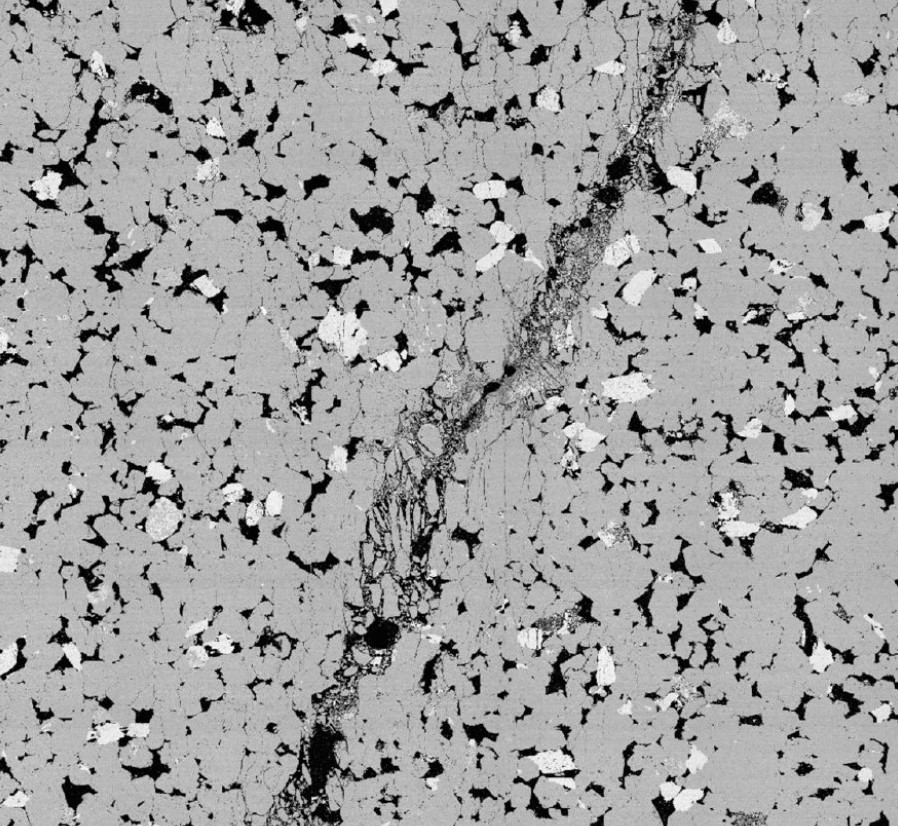}
\caption{\footnotesize{Complex porosity: sandstone}}
\label{fig:c}
\end{subfigure}
 \qquad \qquad\vspace{-0.3cm}
\begin{subfigure}{.35\textwidth}
\includegraphics[scale=0.42]{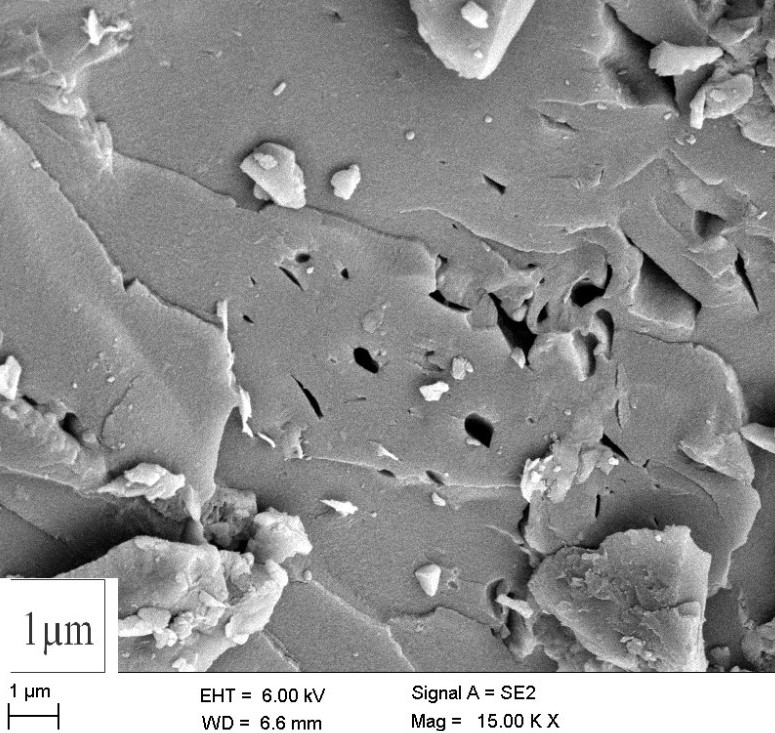}
\caption{\footnotesize{Complex porosity: coal}}
\label{fig:d}
\end{subfigure}
\begin{subfigure}{.35\textwidth}
\includegraphics[scale=0.32]{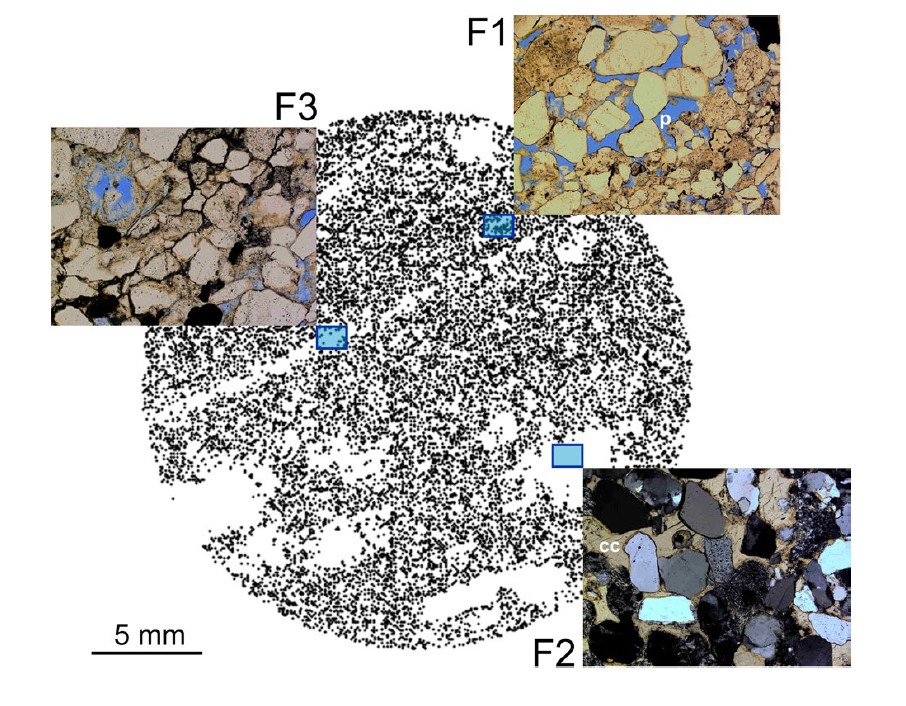}
\caption{\footnotesize{Spatially-distributed sets: sample scale}}
\label{fig:e}
\end{subfigure}
\qquad \qquad
\begin{subfigure}{.4\textwidth}
\includegraphics[scale=0.325]{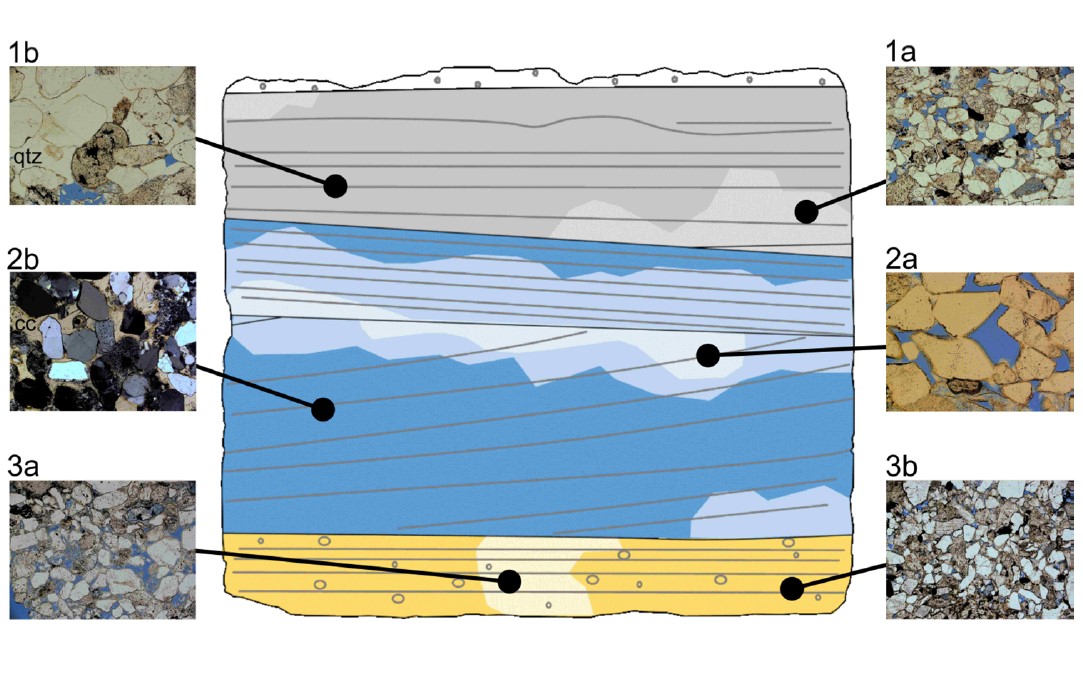}
\caption{\footnotesize{Spatially-distributed sets: large scale}}
\label{fig:f}
\end{subfigure}
\caption{\footnotesize{Literature examples (granted permissions by corresponding authors) with porosity types pertinent to the multi-porous extension. {\bf{a)}} Dolostone with irregular vuggy pores (arrows) and smaller intercrystalline pores from~\citet{WangEtAl15}. {\bf{b)}} Porous sandstone with pores of different sizes from~\citet{FarrellEtAl14}. {\bf{c)}} Fractured sandstone with micropores from~\citet{Rizzo18}. {\bf{d)}} Coal with cleats and micropores from~\citet{PanwarEtAl17}. {\bf{e)}} Sandstone at a sample scale and {\bf{f)}} field scale taken from~\citet{HeidsiekEtAl20}.
} }
\label{fig:two}
\end{figure}

Further, certain anisotropic materials possess multiple pore sets, meaning that a medium may contain porosity clusters that are either weakly connected or isolated from each other. Then, our multi-porous extension of anisotropic poroelasticity theory can be used to describe the mechanics of such media. Although the original Biot theory has a macroscopic nature (quasi-static assumption), in our extension, the mesoscale structures are also distinguished (different pore sets), as schematically depicted in Figure~\ref{fig:one}. Further, one can also search for the micromechanical link to the theory (pore geometry); such linkage is discussed in our parallel paper~\citep{AdamusEtAl23b}.
We subjectively distinguish three porosity types, where our extension is pertinent: hierarchical porosity, complex porosity, and porosity formed by spatially-distributed pore sets. Each type is presented in Figure~\ref{fig:two} and discussed below.

Hierarchical porosity describes a scenario, where there are multiple pore sizes in the medium~\citep{HalliwellEtAl22} forming a nested structure, like in a Russian matryoshka doll~\citep{CowinEtAl09}. Such porosity can be found in geomaterials, extended framework materials, foams, fibres, vascular plants, or body tissues. Examples of hierarchically porous rocks involve carbonates during the dolomitization process (Fig.~\ref{fig:a}) or microporous sandstones having macropores at grain boundaries (Fig.~\ref{fig:b}). The former consist of microporosity, abundant intercrystalline pores, and some irregular vuggy pores that form during the advanced stage of the dolomitization process~\citep{WangEtAl15}. The dolostones are interesting not only from a purely geological perspective (dolomitization) but also from a resource exploration point of view---approximately 50\% of the world's gas and oil reservoirs are in carbonate rocks~\citep{XuEtAl20}, and many ore deposits are hosted in dolostones~\citep{Warren99}. The porosity of clean sandstones is simpler; macroporosity forms a relatively uniform intergranular network, and microporosity forms from detrital and authigenic clays~\citep{ThomsonEtAl19}. Sandstones are also important for petroleum geology since they commonly host oil and gas~\citep[e.g.,][]{BjorlykkeJahren15}. Biomechanical examples involve bone tissue, tendon tissue, or meniscus tissue; extension of the poroelasticity theory can be used to model the mechanical and blood pressure load-driven movements~\citep{CowinEtAl09}. Also, synthesis and applications of hierarchically structured porous materials has become a rapidly evolving field of current interest~\citep{WuEtAl20}. The theory of dual porosity was applied to, for instance, perforated concrete~\citep{CarbajoEtAl17} and sound-absorbing materials~\citep{BecotEtAl08}, like foams and fibres~\citep{LiuEtAl22}.

We refer to a complex porosity where microcracks or larger fractures intersect a medium having background porosity. Such a scenario commonly occurs in geomaterials (e.g., coals, tight-gas sands). Among many fields, the dual-porosity theory is widely applied in petroleum science~\citep[e.g.,][]{LemonnierBourbiaux10, KarimiFardEtAl06, LuoEtAl22}. Originally, the dual-porosity concept was studied to describe the mechanics of conventional fractured reservoirs~\citep{WarrenRoot63}, represented by e.g., fractured porous sandstone (Fig.~\ref{fig:c}). Also, the extended theory of Biot was used to describe unconventional reservoirs, such as coalbed methane~\citep{HuangEtAl22}. 
Coals (Fig.~\ref{fig:d}) contain natural fractures with different permeability than the porous background~\citep{EspinozaEtAl16}. Alternatively, dual porosity can be applied to a fractured geothermal reservoir~\citep{MahzariEtAl21}, such as hydrothermally altered granite~\citep{GenterTraineau92}. 
Note that fractured rocks are usually anisotropic due to the preferred orientation of the fractures and, at a large scale, fracture networks can demonstrate strongly different fluid flow behaviours~\citep[e.g.,][]{Berkowitz02}. Hence, two or more pore sets might need to be considered to describe such phenomena.

We use the term ``spatially-distributed pore sets'' to describe pores that form non-overlapping concentrations. In other words, pores (e.g. cracks) having fixed poroelastic constants are not dispersed throughout the medium, but they constitute a set that is spatially separated from pores having different characteristics (see Fig.~\ref{fig:one}). For instance, \citet{HeidsiekEtAl20} examine a reservoir sandstone with pore sets that vary spatially when viewed at both the sample-scale (Fig.~\ref{fig:e}) and the grid-cell scale relevant to reservoir studies (Fig.~\ref{fig:f}).~\citet{BrantutAben21} measured local fluid-pressure variations within laboratory-scale samples of sandstone and granite. At a large field scale, pore sets and permeabilities also vary spatially. However, to use the multi-porous extension, the considered medium cannot be too large since the solid-matrix stiffness must be uniform, and quasi-static assumptions must be obeyed. To evaluate large spatial domains, like reservoirs, a discretization into smaller scales may be needed~\citep{KarimiFardEtAl06}.

Having discussed each porosity type and indicated studies, where the simplified dual-porosity extension has already been applied, we can now list the anisotropic multi-porous extension potential applications. We believe that the aforementioned extension might be successfully applied to structures having: 
\begin{itemize}
\item{hierarchical porosity with more than two gradations (micropores, mesopores, macropores),}
\item{complex porosity with microcracks or fractures allowing preferential flow (background porosity, microcracks, fractures with slow flow, macro-fractures with fast flow), }
\item{porosity formed by more than two spatially-distributed pore sets,}
\item{mixtures of the porosity types listed above (micropores, macropores, fractures or distributed hierarchical porosity or distributed complex porosity).}
\end{itemize}

%
\newpage
\section{Strain-stress relations}\label{sec3}
Herein, we study the constitutive equations that describe the properties of a deformable elastic medium containing pores.
In these equations, the deformation is viewed at a given instant.
Strains and fluid content changes are related to the changes in stresses and pore pressures through poroelastic constants. 
In other words, all discussed variables (like stress, pore pressure, etc.) are viewed as instantaneous changes, rather than as absolute values.
The deformation is assumed to be small so that linear relations can be utilized.  
In this section, the process of material consolidation or swelling is not considered since time dependency is excluded. We should also note that the sign convention here follows that of elasticity; stresses and strains are positive in the tensile direction. A list of symbols can be found in Appendix A.

Before we describe the constitutive equations, we should define the strains of the porous medium.
The bulk strain tensor,
\begin{equation}\label{strain}
e_{ij}:=\frac{1}{2}\left(\frac{\partial{u_i}}{\partial{x_j}}+\frac{\partial{u_j}}{\partial{x_i}}\right)\,,
\end{equation}
where $u_i$ are the displacements of a skeleton (solid with empty pores) in the $i$-th direction; $x_i$ denote the coordinate axes. Throughout the paper, $i,\,j\in\{1,2,3\}$.
In the case of isotropic medium and hydrostatic confining pressure, the displacements in the principal directions are equal. Hence, the volumetric strain can be written as $e:=\sum_{i}\partial {u_i}/\partial{x_i}=3e_{ii}$.
The change in the fluid content is more difficult to describe due to the varying nature of material porosity. 
Let us use a superscript $(p)$, where $p\in\{1,\dots,\,n\}$, to denote a specific set in a $n$-porosity medium.
Following~\citet{Biot62}, we define 
\begin{equation}\label{fluid}
\zeta^{(p)}:=\phi^{(p)}\sum_{i=1}^3\frac{\partial{u_i}-\partial{U^{(p)}_i}}{\partial{x_i}}\,,
\end{equation}
where $U_i^{(p)}$ is the displacement of the fluid contained in the $p$-th pore set; the volume fraction occupied by such a set is denoted by $\phi^{(p)}\equiv V^{(p)}/V$.
Expression~(\ref{fluid}) describes a relative volumetric strain of a fluid with respect to the solid, loosely referred to as ``fluid increment'' or ``fluid content change'' of a $p$-th pore set. Importantly, the fluid content changes between sets are not taken into account yet. In other words, $\zeta^{(p)}$ should be viewed as a quantity that depends on the external behaviour of the bulk medium considered.

First, let us consider the constitutive equations for isotropic single porosity~\citep{Biot41}. They relate volumetric strain $e$ and fluid increment $\zeta\equiv\zeta^{(1)}$, to the changes in confining pressure $p_c$ and changes in fluid pressure $p_f\equiv p_f^{(1)}$, through drained bulk modulus $K$ and poroelastic coefficients. Changes in pressure are positive in compression.
In the matrix notation,
\begin{equation*}\label{singlepor}
\left[
\begin{array}{c}
e \\
-\zeta \\
\end{array}
\right]
=
\left[
\begin{array}{ccc}
\frac{1}{K} & \frac{1}{3}SB \\
\frac{1}{3}SB & S \\
\end{array}
\right]
\left[
\begin{array}{c}
-p_c \\
-p_f \\
\end{array}
\right]\,,
\end{equation*}
where $S$ and $B$ are the storage and Skempton coefficients, respectively.

Let us consider the constitutive equations for isotropic dual porosity~\citep{BerrymanWang95}. Due to the various characteristic of pore sets, fluid content and pressure changes are not constant throughout the medium. The increments $\zeta^{(1)}$, $\zeta^{(2)}$ are related to the pore pressure changes of the first $p_f^{(1)}$ and second pore set $p_f^{(2)}$ by means of coefficients $a_{ij}$, namely,  
\begin{equation}\label{doublepor1}
\left[
\begin{array}{c}
e \\
-\zeta^{(1)} \\
-\zeta^{(2)} \\
\end{array}
\right]
=
\left[
\begin{array}{ccc}
a_{11} & a_{12} & a_{13}\\
a_{12} & a_{22} & a_{23}\\
a_{13} & a_{23} & a_{33}\\
\end{array}
\right]
\left[
\begin{array}{c}
-p_c \\
-p_f^{(1)} \\
-p_f^{(2)} \\
\end{array}
\right]\,.
\end{equation}
The $a_{ij}$ coefficient matrix must be symmetric because the scalar of the two remaining matrices is the potential energy~\citep{BerrymanWang00}. The existence of the potential energy function implies the invariance of the coefficient matrix under permutations of subscripts $i$ and $j$.
The above form was first given by~\citet{BerrymanWang95}; the original sign convention is preserved.
We can treat the strains and stresses as tensors---instead of scalars---and represent them as $6\times1$ vectors. They are related by the elastic compliances $S_{ijk\ell}$. Hence, without changing any assumptions, we can rewrite isotropic relations~(\ref{doublepor1}) as
\begin{equation*}
\left[
\begin{array}{c}
e_{11} \\
e_{22} \\
e_{33} \\
0 \\
0 \\
0 \\
-\zeta^{(1)} \\
-\zeta^{(2)} \\
\end{array}
\right]
=
\left[
\begin{array}{cccccccc}
S_{1111} & S_{1122} & S_{1122} & 0 & 0 & 0 & -b^{(1)} & -b^{(2)}\\
S_{1122} & S_{1111} & S_{1122} & 0 & 0 & 0 & -b^{(1)} & -b^{(2)}\\
S_{1122} & S_{1122} & S_{1111} & 0 & 0 & 0 & -b^{(1)} & -b^{(2)}\\
0 & 0 & 0 & S_{2323} & 0 & 0 & 0 & 0\\
0 & 0 & 0 & 0 & S_{2323} & 0 & 0 & 0\\
0 & 0 & 0 & 0 & 0 & S_{2323} & 0 & 0\\
-b^{(1)} & -b^{(1)} & -b^{(1)} & 0 & 0 & 0 & a_{22} & a_{23}\\
-b^{(2)} & -b^{(2)} & -b^{(2)} & 0 & 0 & 0 & a_{23} & a_{33}\\
\end{array}
\right]
\left[
\begin{array}{c}
\sigma_{11}\\
\sigma_{22}\\
\sigma_{33}\\
0\\
0\\
0\\
-p_f^{(1)} \\
-p_f^{(2)} \\
\end{array}
\right]\,,
\end{equation*}
where $S_{1122}=S_{1111}-2S_{2323}\,$, $e_{11}=e_{22}=e_{33}=e/3\,$, and $\sigma_{11}=\sigma_{22}=\sigma_{33}=-p_c$\,. Coefficient $b^{(1)}=:-a_{12}/3$ and $b^{(2)}=:-a_{13}/3$\,. This form is analogous to the one shown by~\citet{BerrymanWang00}. Note that pressure changes have the opposite sign to stress changes, $\sigma_{ij}$. 

The equations for isotropic dual porosity can be translated into a general anisotropic case as follows.
\begin{equation}\label{doubleporaniso}
\left[
\begin{array}{c}
e_{11} \\
e_{22} \\
e_{33} \\
e_{23} \\
e_{13} \\
e_{12} \\
\zeta^{(1)} \\
\zeta^{(2)} \\
\end{array}
\right]
=
\left[
\begin{array}{cccccccc}
S_{1111} & S_{1122} & S_{1133} & S_{1123} & S_{1113} & S_{1112} & b_{11}^{(1)} & b_{11}^{(2)}\\
S_{1122} & S_{2222} & S_{2233} & S_{2223} & S_{2213} & S_{2212} & b_{22}^{(1)} & b_{22}^{(2)}\\
S_{1133} & S_{2233} & S_{3333} & S_{3323} & S_{3313}  & S_{3312} & b_{33}^{(1)} & b_{33}^{(2)}\\
S_{1123}  & S_{2223}  & S_{3323}  & 2S_{2323} & 2S_{2313} & 2S_{2312} & b_{23}^{(1)} & b_{23}^{(2)}\\
S_{1113}  & S_{2213}  & S_{3313}  & 2S_{2313} & 2S_{1313} & 2S_{1312} & b_{13}^{(1)} & b_{13}^{(2)}\\
S_{1112}  & S_{2212}  & S_{3312}  & 2S_{2312}  & 2S_{1312}  & 2S_{1212} & b_{12}^{(1)} & b_{12}^{(2)}\\
b_{11}^{(1)} & b_{22}^{(1)} & b_{33}^{(1)} & b_{23}^{(1)} & b_{13}^{(1)} & b_{12}^{(1)} & a_{22} & a_{23}\\
b_{11}^{(2)} & b_{22}^{(2)} & b_{33}^{(2)} & b_{23}^{(2)} & b_{13}^{(2)} & b_{12}^{(2)} & a_{23} & a_{33}\\
\end{array}
\right]
\left[
\begin{array}{c}
\sigma_{11}\\
\sigma_{22}\\
\sigma_{33}\\
\sigma_{23}\\
\sigma_{13}\\
\sigma_{12}\\
p_f^{(1)} \\
p_f^{(2)} \\
\end{array}
\right]\,.
\end{equation}
We see that coefficients $b^{(p)}$ transformed into second-rank tensors, and negative signs have cancelled. Factor $2$ in front of certain compliances appeared due to the index symmetries of the stress tensor. The meaning of $a_{ij}$ and $b_{ij}^{(p)}$ will be explained in the next section. Straightforwardly, we generalise the above form to the multiple-porosity, $n$-set scenario, namely, 
\begin{equation}\label{multiporaniso}
\left[
\begin{array}{c}
\bf{e} \\
\zeta^{(1)} \\
\vdots \\
\zeta^{(n)} \\
\end{array}
\right]
=
\left[
\begin{array}{cccc}
\bf{S} & \bf{b^{(1)}} & \cdots & \bf{b^{(n)}} \\
{\bf{b^{(1)}}}^T  & a_{2,2}  &\cdots & a_{2,n+1} \\
\vdots   & \vdots   & \ddots  & \vdots \\
{\bf{b^{(n)}}}^T  & a_{2, n+1}  & \cdots  & a_{n+1,n+1} \\
\end{array}
\right]
\left[
\begin{array}{c}
\bf{\sigma}\\
p_f^{(1)} \\
\vdots \\
p_f^{(n)} \\
\end{array}
\right]\,,
\end{equation}
where $\bf{e}$, $\bf{b^{(p)}}$, $\bf{\sigma}$ are $6\times1$ vectors, $\bf{S}$ is a $6\times6$ matrix and $^T$ denotes the transposition. In the next sections, we will strive to get more insight into the novel equations~(\ref{doubleporaniso})--(\ref{multiporaniso}) given above.

\section{Determination of $a_{ij}$ and $b_{ij}^{(p)}$}\label{sec4}
In their present form, $a_{ij}$ and $b_{ij}^{(p)}$ are difficult to measure in practice, and their physical meaning is unclear. Therefore, in this section, we describe them in terms of elastic compliances and poroelastic coefficients (Skempton-like and storages). 
We consider various poroelastic boundary conditions that lead to the $a_{ij}$ and $b_{ij}^{(p)}$ determination. This way, a few universal constraints for weakly-connected sets are also provided. 

The standard boundary conditions of poroelasticity are referred to as the drained and undrained states, defined by no change in fluid pressure and no change in fluid content, respectively. 
Another possible limit is that of constant confining stress, $\bf{\sigma}=\bf{0}$ (this is not an absolute value). Such a condition is strived to be achieved in the fluid injection tests. However, in the case of a multiset extension, more variables require more boundary conditions that need to be considered. Besides, if sets are weakly connected, the long-time limit leads to pressure equilibration. In turn, the constraints for weakly-connected sets can be formulated. Below, we invoke those scenarios (test types) that provide the aforementioned constraints and useful information on $a_{ij}$ and $b_{ij}^{(p)}$. For simplicity, we assume dual porosity (two pore sets) that can be easily extended to $n>2$ considerations.
\subsection{Drained test, long-time limit}\label{sec:drained}
In this test type, pressure throughout the pores is in equilibrium and constant. In other words, $p_f^{(1)}=p_f^{(2)}=0$. Hence, strains from equation~(\ref{doubleporaniso}) are simplified to $\bf{e}=\bf{S\sigma}$. In such drained conditions, the elastic compliances ($S_{ijk\ell}$) can be determined. In the case of isotropy, we get $a_{11}=1/K$, where $K$ is the drained bulk modulus.
\subsection{Undrained test, long-time limit}\label{sec:un_long}
This scenario provides us with the first constraint of the theory, assuming that weak connections exist between sets. If a rock sample has a closed system---so that the total fluid content change, $\zeta_{tot}$, is zero---we get,
\begin{equation*}
\zeta_{tot}:=\zeta^{(1)}+\zeta^{(2)}=0\,,
\end{equation*}
and after a sufficiently long time,
\begin{equation*}
p_f^{(1)}=p_f^{(2)}=p_f\,.
\end{equation*}
In this way, we obtain the following constraints
\begin{equation}\label{constraint1}
e_{ij}=\sum_{k=1}^3\sum_{\ell=1}^3S_{ijk\ell}\sigma_{k\ell}+\left(b_{ij}^{(1)}+b_{ij}^{(2)}\right)p_f\,
\end{equation}
and
\begin{equation}\label{constraint2}
p_f=-\frac{1}{a_{22}+2a_{23}+a_{33}}\sum_{k=1}^3\sum_{\ell=1}^3\left(b_{k\ell}^{(1)}+b_{k\ell}^{(2)}\right)\sigma_{k\ell}\,.
\end{equation}
These are valid only if the sets are not isolated (pressure equilibration over time is possible). In the case of uniaxial stress, we can derive bulk Skempton coefficients from (\ref{constraint2}),
\begin{equation}\label{skempton}
\left.-\frac{p_f}{\sigma_{ij}}\right|_{\zeta_{tot}=0}\equiv \frac{1}{3}B_{k\ell}=\frac{b_{k\ell}^{(1)}+b_{k\ell}^{(2)}}{a_{22}+2a_{23}+a_{33}}\,,
\end{equation}
and obtain the undrained elastic compliances $S_{ijk\ell}^u$ from~(\ref{constraint1}),
\begin{equation}\label{undrained}
\left.\frac{e_{ij}}{\sigma_{k\ell}}\right|_{\zeta_{tot}=0}\equiv S^{u}_{ijk\ell}=S_{ijk\ell}-\frac{1}{3}\left(b_{ij}^{(1)}+b_{ij}^{(2)}\right)B_{k\ell}\,.
\end{equation}
Note that due to the pressure equilibration, the aforementioned bulk Skempton coefficients are equivalent to the Skempton coefficients of a single-porosity system.
\subsection{Fluid injection test, long-time limit}\label{sec:fluid_long}
Another constraint may be derived if we perform a standard test of fluid injection, where the applied stress is constant ($\bf{\sigma}=\bf{0}$). We can again assume that the pore pressure equilibrates throughout the medium. This way, we may get the third (and last) long-time constraint
\begin{equation}\label{constraint3}
\left.\frac{\zeta_{tot}}{p_f}\right|_{\bf{\sigma}=\bf{0}}\equiv S=a_{22}+2a_{23}+a_{33}\,
\end{equation}
that describes the storage coefficient of the bulk medium equivalent to the storage of a single-porosity system. Note that upon combining~(\ref{skempton})--(\ref{constraint3}), the undrained compliances are
\begin{equation}\label{undrained2}
S^{u}_{ijk\ell}=S_{ijk\ell}+\Delta_{ijk\ell}:=S_{ijk\ell}-\frac{1}{9}SB_{ij}B_{k\ell}\,,
\end{equation}
where $\Delta_{ijk\ell}$ is defined as the difference between the undrained and drained compliance tensors; hence, it accounts for the effect of fluids.
\subsection{Fluid injection test, drained first set}\label{sec:fluid_drained1}
Consider another scenario where again, the applied stress is constant $\bf{\sigma}=\bf{0}$. Let the fluid be injected directly into the first set (e.g., background porosity) so that upon a longer time, it becomes drained, $p_f^{(1)}=0$, since fluid migrated towards the second set. If sets are weakly connected, it may happen that the pressure in the second set is---on average---still changing due to the fluid outflow. On the other hand, the change of fluid content is almost null at the connection points between sets. Hence, we consider a period when pore pressures in a medium are not equilibrated yet, $p_f^{(2)}\neq 0$. 

Another experiment can lead to $\bf{\sigma}=\bf{0}$, $p_f^{(1)}=0$, and $p_f^{(2)}\neq 0$ if the fluid was injected in the second set, and in a short period, it did not have enough time to migrate towards another set. This way, the first set could remain dry (or drained). 

\begin{figure}[!htbp]
\centering
\begin{subfigure}{.45\textwidth}
\includegraphics[scale=0.5]{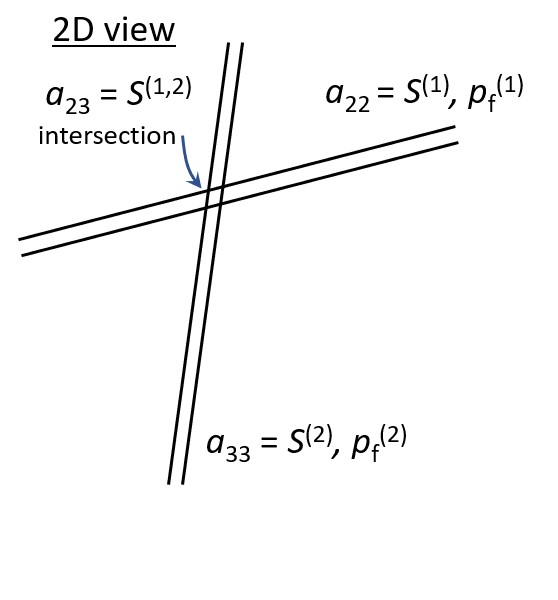}
\caption{\footnotesize{ }}
\label{fig:e}
\end{subfigure}
\begin{subfigure}{.45\textwidth}
\includegraphics[scale=0.405]{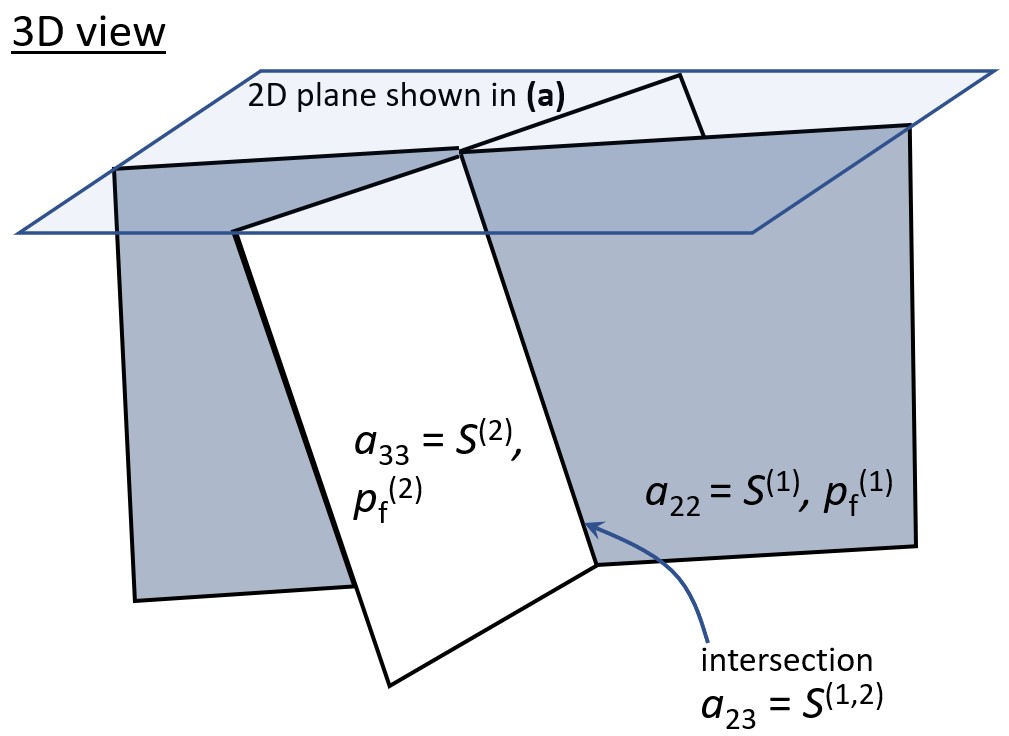}
\caption{\footnotesize{}}
\label{fig:f}
\end{subfigure}
\caption{\footnotesize{Schematic illustration showing different pore sets and how the terms in equations~(\ref{a23})--(\ref{a33}) and (\ref{a22}) relate to physical objects. The coefficient $a_{23}$ is the cross-term describing the storage coefficient of the intersections of set $1$ and set $2$. A mapping of this term to the example of two intersecting fracture sets is shown in 2D {\bf{(a)}} and 3D {\bf{(b)}}.
} }
\label{fig:a23}
\end{figure}

In the case of either experiment, we get
\begin{align*}
e_{ij}&=b_{ij}^{(2)}p_f^{(2)}\,,\\
\zeta^{(1)}&=a_{23}p_f^{(2)}\,,\\
\zeta^{(2)}&=a_{33}p_f^{(2)}\,.
\end{align*}
Thus, we can define
\begin{align}\label{coeff123}
\left.\frac{e_{ij}}{p_f^{(2)}}\right|_{{\bf{\sigma}=\bf{0}},\,p_f^{(1)}=0}&:=b_{ij}^{(2)}\,,\\\label{a23}
\left.\frac{\zeta^{(1)}}{p_f^{(2)}}\right|_{{\bf{\sigma}=\bf{0}},\,p_f^{(1)}=0}&\equiv S^{(1,2)}:=a_{23}\,,\\ \label{a33}
\left.\frac{\zeta^{(2)}}{p_f^{(2)}}\right|_{{\bf{\sigma}=\bf{0}},\,p_f^{(1)}=0}&\equiv S^{(2)}:=a_{33}\,, 
\end{align}
where $b_{ij}^{(2)}$ is viewed as a poroelastic expansion due to fluids in the second set, scalar $S^{(1,2)}$ is the storage coefficients of weak connections between both sets, and $S^{(2)}$ is the storage coefficients of the second set. If sets are isolated, $a_{23}=0$. It makes sense that isolated sets lead to $S^{(1,2)}=0$ since there are no connections where the fluid could be stored. In Figure~\ref{fig:a23}, we illustrate the physical meaning of $a_{ij}$ by considering sets $1$ and $2$ as two independent fractures with storages $S^{(1)}=a_{22}$, and $S^{(2)}=a_{33}$, respectively. Their connection would correspond to their intersection line, which holds its own storage $S^{(1,2)}=a_{23}$.
\subsection{Fluid injection test, drained second set}\label{sec:fluid_drained2}
If the set numbers are interchanged, then an analogous fluid injection experiments can be performed, where $\bf{\sigma}=\bf{0}$, $p_f^{(2)}=0$, and $p_f^{(1)}\neq 0$. 
We obtain,
\begin{align*}
e_{ij}&=b_{ij}^{(1)}p_f^{(1)}\,,\\
\zeta^{(1)}&=a_{22}p_f^{(1)}\,,\\
\zeta^{(2)}&=a_{23}p_f^{(1)}\,
\end{align*}
and define
\begin{align}
\left.\frac{e_{ij}}{p_f^{(1)}}\right|_{{\bf{\sigma}=\bf{0}},\,p_f^{(2)}=0}&:=b_{ij}^{(1)}\,,\\ \label{a22}
\left.\frac{\zeta^{(1)}}{p_f^{(1)}}\right|_{{\bf{\sigma}=\bf{0}},\,p_f^{(2)}=0}&\equiv S^{(1)}:=a_{22}\,,\\  \label{coeff321}
\left.\frac{\zeta^{(2)}}{p_f^{(1)}}\right|_{{\bf{\sigma}=\bf{0}},\,p_f^{(2)}=0}&\equiv S^{(2,1)}:=a_{23}=S^{(1,2)}\,,
\end{align}
where $b_{ij}^{(1)}$ is the poroelastic expansion due to fluids in the first set, and $S^{(1)}$ is the storage coefficient of this set (see Figure~\ref{fig:a23}).
\subsection{Undrained first set, drained second set}\label{sec:ud1}
Consider an abrupt change of stress imposed on a jacketed sample. 
Assume that fluid outflows from the second set due to the insertion of a tube ($p_f^{(2)}=0$), whereas the first set remains approximately closed ($\zeta^{(1)}=0$). Such a short-time experiment can work only if there is very little or no flow between sets---that is also required by the theory extension.
We obtain,
\begin{align*}
e_{ij}&=\sum_{k=1}^3\sum_{\ell=1}^3S_{ijk\ell}\sigma_{k\ell}+b_{ij}^{(1)}p_f^{(1)}\,,\\
0&=\sum^3_{i=1}\sum^3_{j=1}b_{ij}^{(1)}\sigma_{ij}+a_{22}p_f^{(1)}\,,\\
\zeta^{(2)}&=\sum^3_{i=1}\sum^3_{j=1}b_{ij}^{(2)}\sigma_{ij}+a_{23}p_f^{(1)}\,.
\end{align*}
Assuming the uniaxial stress, we can define
\begin{equation}\label{SkemptonUndrained1}
\left.-\frac{p_f^{(1)}}{\sigma_{ij}}\right|_{\zeta^{(1)}=0,\,p_f^{(2)}=0}\equiv \frac{1}{3}B_{ij}^{(1)}:=\frac{b_{ij}^{(1)}}{a_{22}}\,,\qquad
\left.\frac{e_{ij}}{\sigma_{k\ell}}\right|_{\zeta^{(1)}=0,\,p_f^{(2)}=0}\equiv S_{ijk\ell}^{u\,\,(1)}:=S_{ijk\ell}-\frac{1}{3}b_{ij}^{(1)}B_{k\ell}^{(1)}\,,
\end{equation}
where $B_{ij}^{(1)}$ is the Skempton-like tensor of the first set and $S_{ijk\ell}^{u\,\,(1)}$ is the undrained compliance tensor of the first set. 
Now the meaning of $b_{ij}^{(1)}$ becomes more tangible since knowing the definition of $a_{22}$ we can write
\begin{equation}\label{b_alt1}
b_{ij}^{(1)}:=\frac{1}{3}S^{(1)}B_{ij}^{(1)}\,.
\end{equation}
Having~(\ref{b_alt1}), we notice that definition of $S_{ijk\ell}^{u\,\,(1)}$ is analogous to the meaning of $S_{ijk\ell}^{u}$ from~(\ref{undrained2}), namely,
\begin{equation*}\label{fluideffect1}
S^{u\,\,(1)}_{ijk\ell}=S_{ijk\ell}+\Delta_{ijk\ell}^{(1)}:=S_{ijk\ell}-\frac{1}{9}S^{(1)}B^{(1)}_{ij}B^{(1)}_{k\ell}\,,
\end{equation*}
where $\Delta^{(1)}_{ijk\ell}$ accounts for the effect of fluid caused by the first set.
\subsection{Undrained second set, drained first set}\label{sec:ud2}
In the analogous test, the order of sets is switched. This way, $p_f^{(1)}=0$ and $\zeta^{(2)}=0$. 
In the case of uniaxial stress, we get 
\begin{equation}\label{SkemptonUndrained2}
\left.-\frac{p_f^{(2)}}{\sigma_{ij}}\right|_{\zeta^{(2)}=0,\,p_f^{(1)}=0}\equiv \frac{1}{3}B_{ij}^{(2)}:=\frac{b_{ij}^{(2)}}{a_{33}}
\,,\qquad
\left.\frac{e_{ij}}{\sigma_{k\ell}}\right|_{\zeta^{(1)}=0,\,p_f^{(2)}=0}\equiv S_{ijk\ell}^{u\,\,(1)}:=S_{ijk\ell}-\frac{1}{3}b_{ij}^{(1)}B_{k\ell}^{(1)}
\end{equation}
and
\begin{equation}\label{b_alt2}
b_{ij}^{(2)}:=\frac{1}{3}S^{(2)}B_{ij}^{(2)}\,,
\end{equation}
\begin{equation*}
S^{u\,\,(2)}_{ijk\ell}=S_{ijk\ell}+\Delta_{ijk\ell}^{(2)}:=S_{ijk\ell}-\frac{1}{9}S^{(2)}B^{(2)}_{ij}B^{(2)}_{k\ell}\,.
\end{equation*}
\subsection{Summary of conditions}
Let us discuss each scenario (test type), where different boundary conditions were assumed. First, we point out three test types that require a long-time limit.
Naturally, a drained, long-time test type is indispensable to determine compliances, $S_{ijk\ell}$. Further, two scenarios from Sections~\ref{sec:un_long} and~\ref{sec:fluid_long} are important to establish the constraints~(\ref{constraint1}),~(\ref{constraint2}), and~(\ref{constraint3}). They are necessary only if sets are not perfectly isolated. In other words, in the case of isolated sets, we need one test type only (drained, long-time). Note that the amount of required test types is not affected by the number of sets embedded in the solid matrix. Boundary conditions from Sections~\ref{sec:drained}--\ref{sec:fluid_long} can be straightforwardly translated into $n>2$ considerations without the necessity of introducing new scenarios.

Further, let us indicate those scenarios that determine $a_{ij}$ and $b_{ij}^{(p)}$, and where a long-time limit is not required.
In the case of dual porosity, to determine $a_{ij}$ and $b_{ij}^{(p)}$, we need only two types of tests. A reader can choose to either perform the fluid injection tests presented in Sections~\ref{sec:fluid_drained1}--\ref{sec:fluid_drained2} or perform the ``drained-undrained'' tests from Sections~\ref{sec:ud1}--\ref{sec:ud2}. In the first possibility, $a_{ij}$ (defined as storages) and $b_{ij}^{(p)}$ (defined as poroelastic expansions) are directly measured~(\ref{coeff123})--(\ref{coeff321}). In the second possibility, a combination of measured Skempton-like tensors and undrained compliances lead to indirect determination of $a_{ij}$ and $b_{ij}^{(p)}$. Specifically, 
equation~(\ref{SkemptonUndrained1}) gives $a_{22}$ and $b_{ij}^{(1)}$, equation~(\ref{SkemptonUndrained2}) provides $a_{33}$ and $b_{ij}^{(2)}$, and additional equation~(\ref{constraint3}) from fluid-injection long-time test gives remaining $a_{23}$.
In the case of multiple pore sets, we require additional scenarios to determine all $\{a_{ij}^{(1)},\, b_{ij}^{(1)}\},\,\dots,\,\{a_{ij}^{(n)},\, b_{ij}^{(n)}\}$. In the fluid injection tests, all sets except one must be drained ($n$ combinations in total). The alternative tests that determine Skempton-like tensors and undrained compliances also require the sets to be drained except for one that needs to be undrained (again $n$ combinations in total). The exact amount of test types can be easily deduced and is listed in Table~\ref{tab:plan}.
\begin{table}[!htbp]
\begin{tabular}{cc}
 pressure equilibration      &   required test types  \\
\cmidrule{1-2}
possible (weak connections)    &    $1^*+\dot{2}+\hat{n}$    \\          
impossible (all sets isolated)   &      $1^*+\hat{n}$       \\
\cmidrule{1-2}
\end{tabular}
\caption{\small{Number of boundary test types for $n$-set extension. The required number determines drained compliances (asterix), constraints coming from long-time pressure equilibration (dot), and coefficients $a_{ij}$ and $b^{(p)}_{ij}$ (hat). Poroelastic coefficients can be determined directly (fluid-injection tests) or indirectly (``drained-undrained'' tests).}}
\label{tab:plan}
\end{table}
%
\section{Alternative formulations of strain-stress relations}\label{sec5}
It might be beneficial to reformulate strain-stress relations by describing $a_{ij}$ and $b_{ij}^{(p)}$ in terms of storages and Skempton-like coefficients, as shown in~(\ref{a23})-(\ref{a33}), (\ref{a22}), (\ref{b_alt1}), and (\ref{b_alt2}). We can rewrite equation~(\ref{doubleporaniso}) in the tensorial notation as,
\begin{align*}
e_{ij}&=\sum^3_{k=1}\sum^3_{\ell=1}S_{ijk\ell}\sigma_{k\ell}+\frac{1}{3}S^{(1)}B_{ij}^{(1)}p_f^{(1)}+\frac{1}{3}S^{(2)}B_{ij}^{(2)}p_f^{(2)}\,,\\
\zeta^{(1)}&=\frac{1}{3}S^{(1)}\sum^3_{k=1}\sum^3_{\ell=1}B^{(1)}_{k\ell}\sigma_{k\ell}+S^{(1)}p_f^{(1)}+S^{(1,2)}p_f^{(2)}\,,\\
\zeta^{(2)}&=\frac{1}{3}S^{(2)}\sum^3_{k=1}\sum^3_{\ell=1}B^{(2)}_{k\ell}\sigma_{k\ell}+S^{(1,2)}p_f^{(1)}+S^{(2)}p_f^{(2)}\,,
\end{align*}
where in the idealized case of perfectly isolated sets, $S^{(1,2)}=0$.
Having storage and Skempton-like coefficients that describe each set, we rewrite multi-porous equation~(\ref{multiporaniso}) as
\begin{align}\label{version2a}
e_{ij}&=\sum^3_{k=1}\sum^3_{\ell=1}S_{ijk\ell}\sigma_{k\ell}+\sum_{p=1}^n\frac{1}{3}S^{(p)}B_{ij}^{(p)}p_f^{(p)}\,,\\ \label{version2b}
\zeta^{(p)}&=\frac{1}{3}S^{(p)}\sum^3_{k=1}\sum^3_{\ell=1}B^{(p)}_{k\ell}\sigma_{k\ell}+S^{(p)}p_f^{(p)}+\sum_{q=1}^nS^{(p,q)}p_f^{(q)}\,,
\end{align}
where $q\in\mathbb{N},\,1\leq q\leq n$, and $q\neq p$. If certain set $p$ is isolated from $q$, then $S^{(p,q)}=0$. The equations for all isolated sets are given and discussed explicitly in our parallel paper~\citep{AdamusEtAl23b}.

So far, we have presented two versions of strain-stress relations for a multiple-porosity system. Equations~(\ref{multiporaniso}) used coefficients $a_{ij}^{(p)}$ and $b_{ij}^{(p)}$, whereas equations~(\ref{version2a})--(\ref{version2b}) utilized more tangible Skempton-like and storage coefficients. Below, we introduce new parameters that facilitate the alternative description of equations relating strains and stresses useful for further, time-dependent analysis. 
After algebraic manipulations on equations~(\ref{version2a})--(\ref{version2b}), we obtain mixed strain-stress formulations,
\begin{align}\label{version3}
\sigma_{ij}&=\sum^3_{k=1}\sum^3_{\ell=1}C_{ijk\ell}e_{k\ell}-\sum_{p=1}^n\alpha_{ij}^{(p)}p_f^{(p)}\,,\\ \label{mixedzeta}
\zeta^{(p)}&=\sum^3_{k=1}\sum^3_{\ell=1}\alpha_{k\ell}^{(p)}e_{k\ell}+\frac{1}{M^{(p)}}p_f^{(p)}+\sum_{q=1}^n\frac{1}{M^{(p,q)}}p_f^{(q)}\,,
\end{align}
where again $q\in\mathbb{N},\,1\leq q\leq n$, $q\neq p$, and $C_{ijk\ell}$ denotes elasticity tensor.
Further,
\begin{equation}\label{biotwillis}
\alpha_{ij}^{(p)}:=\frac{1}{3}S^{(p)}\sum_{k=1}^3\sum_{\ell=1}^3C_{ijk\ell}B^{(p)}_{k\ell}
\end{equation}
is the Biot-like tensor for each set of pores and 
\begin{equation}\label{M}
\frac{1}{M^{(p)}}:=S^{(p)}-\sum^3_{i=1}\sum^3_{j=1}\sum^3_{k=1}\sum^3_{\ell=1}S_{ijk\ell}\alpha^{(p)}_{ij}\alpha_{k\ell}^{(p)}\,,
\end{equation}
\begin{equation*}\label{M2}
\frac{1}{M^{(p,q)}}:=S^{(p,q)}-\sum^3_{i=1}\sum^3_{j=1}\sum^3_{k=1}\sum^3_{\ell=1}S_{ijk\ell}\alpha^{(p)}_{ij}\alpha_{k\ell}^{(q)}
\end{equation*}
describe fluid storage under no frame deformation. To grasp the physical meaning of the above-mentioned coefficients, let us think of fluid injection tests analogous to the ones from Section~\ref{sec:fluid_drained1}--\ref{sec:fluid_drained2}, where $\bf{e}=\bf{0}$ instead of $\bf{\sigma}=\bf{0}$ is required. Then, we obtain
\begin{align*}
\left.\frac{\sigma_{ij}}{p_f^{(2)}}\right|_{{\bf{e}=\bf{0}},\,p_f^{(1)}=0}&\equiv -\alpha_{ij}^{(2)}\,,\\
\left.\frac{\zeta^{(1)}}{p_f^{(2)}}\right|_{{\bf{e}=\bf{0}},\,p_f^{(1)}=0}&\equiv \frac{1}{M^{(1,2)}}\,,\\
\left.\frac{\zeta^{(2)}}{p_f^{(2)}}\right|_{{\bf{e}=\bf{0}},\,p_f^{(1)}=0}&\equiv \frac{1}{M^{(2)}}\,
\end{align*}
and
\begin{align*}
\left.\frac{\sigma_{ij}}{p_f^{(1)}}\right|_{{\bf{e}=\bf{0}},\,p_f^{(2)}=0}&\equiv -\alpha_{ij}^{(2)}\,,\\
\left.\frac{\zeta^{(1)}}{p_f^{(1)}}\right|_{{\bf{e}=\bf{0}},\,p_f^{(2)}=0}&\equiv \frac{1}{M^{(1)}}\,,\\
\left.\frac{\zeta^{(2)}}{p_f^{(1)}}\right|_{{\bf{e}=\bf{0}},\,p_f^{(2)}=0}&\equiv \frac{1}{M^{(2,1)}}=\frac{1}{M^{(1,2)}}\,, 
\end{align*}
respectively.
Note that definitions~(\ref{biotwillis}) and~(\ref{M}) are analogous to the~\citet{Cheng97} definitions for single porosity (his equations (20) and (22)).
The mixed formulation will be explicitly used in the consolidation equations in the next section.
Also, one can try to introduce stress-strain relations by switching $\zeta^{(p)}$ with $p_f^{(p)}$ and reformulating poroelastic coefficients~\citep{Mehrabian18}. However, in contrast to the original Biot theory, such a formulation may become complicated due to the existence of multiple pore pressures. 
In a simpler case of isolated sets, we obtain
\begin{align}\label{version4}
\sigma_{ij}&=\sum^3_{i=1}\sum^3_{j=1}C^u_{ijk\ell}e_{k\ell}-\sum_{p=1}^nM^{(p)}\alpha_{ij}^{(p)}\zeta^{(p)}\,,\\ \label{version4b}
p_f^{(p)}&=-M^{(p)}\sum^3_{k=1}\sum^3_{\ell=1}\alpha^{(p)}_{k\ell}e_{k\ell}+M^{(p)}\zeta^{(p)}\,,
\end{align}
where the undrained elasticity parameters
\begin{equation*}\label{C}
C_{ijk\ell}^u:=C_{ijk\ell}+\sum_{p=1}^nM^{(p)}\alpha^{(p)}_{ij}\alpha^{(p)}_{k\ell}\,.
\end{equation*}
Stress-strain relations for a more complicated case of weakly-connected sets are given in Appendix~\ref{ap1}.
\section{Governing equations of consolidation}\label{sec6}
Herein, we study the governing equations that describe the transient phenomenon of three-dimensional consolidations. These are differential equations governing the distributions of stress and fluid content change and settlement as a function of time in a medium under given loads. If a fluid content is increased due to an increase in the volume of pores, the governing equations describe the opposite process of swelling instead of consolidation. For convenience, we derive the case of dual porosity. Nevertheless, at the end of the section, the multi-porous generalisation is additionally presented.

To obtain the governing equations, first, changes in externally applied stresses must satisfy the equilibrium conditions; namely,
\begin{equation}\label{equilibrium}
\sum_{j=1}^3\frac{\partial{\sigma_{ij}}}{\partial{x_j}}=0\,.
\end{equation}
We assume that body forces can be neglected.
Then, we insert~(\ref{version3}) into~(\ref{equilibrium}) and rewrite strains in terms of displacements~(\ref{strain}).
We obtain three governing equations
\begin{equation}\label{governing}
\sum_{j=1}^3\left[\,\,\sum_{k=1}^3\sum_{\ell=1}^3\frac{1}{2}C_{ijk\ell}\left(\frac{\partial u_{k}}{\partial {x_j}\partial {x_\ell}}+\frac{\partial u_{\ell}}{\partial {x_j}\partial {x_k}}\right)-\alpha^{(1)}_{ij}\frac{\partial{p_f^{(1)}}}{\partial{x_j}}-\alpha^{(2)}_{ij}\frac{\partial{p_f^{(2)}}}{\partial{x_j}}\right]=0\,.
\end{equation}
To obtain the rest of the governing equations, let us define the flux of the fluid through pore set $p$ as
\begin{equation}\label{q}
q_i^{(p)}:=\phi^{(p)}\left(\frac{\partial{U_i^{(p)}}}{\partial{t}}-\frac{\partial{u_i}}{\partial{t}}\right)\,.
\end{equation}
Having defined the fluid content changes~(\ref{fluid}) and the fluid flux~(\ref{q}), we can formulate the equations of fluid continuity. Assuming that the fluid is incompressible and that a certain amount diffuses internally, 
\begin{align}\label{mass}
\sum_{i=1}^3\frac{\partial{q_i^{(1)}}}{\partial{x_i}}+\frac{\partial{\zeta^{(1)}}}{\partial{t}}+\frac{\partial{\zeta^{(1,2)}}}{\partial{t}}=0\,,\\ \label{mass2}
\sum_{i=1}^3\frac{\partial{q_i^{(2)}}}{\partial{x_i}}+\frac{\partial{\zeta^{(2)}}}{\partial{t}}+\frac{\partial{\zeta^{(2,1)}}}{\partial{t}}=0\,,
\end{align}
where the interset fluid flow
\begin{equation*}
\frac{\partial{\zeta^{(p,q)}}}{\partial{t}}:=\Gamma^{(p,q)}\left(p_f^{(p)}-p_f^{(q)}\right)
\end{equation*} 
is assumed to be proportional to the difference in pore pressures.
Equations~(\ref{mass})--(\ref{mass2}) state that the external change in the amount of fluid in a pore set at any instant is balanced by the fluid flowing into the set whilst a certain amount of fluid is diffused inside the medium. It is required that $\partial{\zeta^{(1,2)}}/\partial{t}=-\partial{\zeta^{(2,1)}}/\partial{t}$ since the internal diffusion between the sets must be equal; fluid volume is assumed to be conserved and not compressed. Note that in the case of $n>2$ sets, there is more than a single leakage constant $\Gamma$. For instance, if $n=4$, we get six possibilities, namely, $\Gamma^{(1,2)}$, $\Gamma^{(1,3)}$, $\Gamma^{(1,4)}$, $\Gamma^{(2,3)}$, $\Gamma^{(2,4)}$, and $\Gamma^{(3,4)}$. In total, there are $\sum_{p=1}^n(n-p)$ leakage constants.

Now, let us invoke a crucial constitutive relation. Darcy's law for dual porosity can be written as
\begin{align}\label{darcy}
q_i^{(1)}=-\sum_{j=1}^3 \frac{k_{ij}^{(1)}}{\mu}\frac{\partial {p_f^{(1)}}}{\partial{x_j}}-\sum_{j=1}^3 \frac{k_{ij}^{(1,2)}}{\mu}\frac{\partial {p_f^{(2)}}}{\partial{x_j}}\,,\\\label{darcy2}
q_i^{(2)}=-\sum_{j=1}^3 \frac{k_{ij}^{(2)}}{\mu}\frac{\partial {p_f^{(2)}}}{\partial{x_j}}-\sum_{j=1}^3 \frac{k_{ij}^{(2,1)}}{\mu}\frac{\partial {p_f^{(1)}}}{\partial{x_j}}\,,
\end{align}
where $\mu$ denotes viscosity. Fluid is assumed to be of the same type throughout the medium.
Note that~(\ref{darcy})--(\ref{darcy2}) are the anisotropic extensions of~\citet{Aifantis80a} expressions. 
We can combine the aforementioned Darcy's law~(\ref{darcy})--(\ref{darcy2}) with the continuity equations~(\ref{mass})--(\ref{mass2}) and fluid content changes from strain-stress relations~(\ref{mixedzeta}). This way, we obtain the remaining set of governing equations, namely,
\begin{equation}\label{governing2a}
\begin{aligned}
\sum_{i=1}^3&\sum_{j=1}^3\left( \frac{k_{ij}^{(1)}}{\mu}\frac{\partial^2 {p_f^{(1)}}}{\partial{x_i}\partial{x_j}}+ \frac{k_{ij}^{(1,2)}}{\mu}\frac{\partial^2 {p_f^{(2)}}}{\partial{x_i}\partial{x_j}}\right)+\Gamma\left(p_f^{(2)}-p_f^{(1)}\right)
=\hphantom{XXXXXXXXXXXXX} \\
&\hphantom{XXXXXXXXXXXXX}\frac{1}{2}\sum^3_{i=1}\sum^3_{j=1}\alpha_{ij}^{(1)}\frac{\partial}{\partial{t}}\left(\frac{\partial{u_{i}}}{\partial{x_j}}+\frac{\partial{u_{j}}}{\partial{x_i}}\right)+\frac{1}{M^{(1)}}\frac{\partial{p_f^{(1)}}}{\partial{t}}+\frac{1}{M^{(1,2)}}\frac{\partial{p_f^{(2)}}}{\partial{t}}\,,
\end{aligned}
\end{equation} 
\begin{equation}\label{governing2b}
\begin{aligned}
\sum_{i=1}^3&\sum_{j=1}^3\left( \frac{k_{ij}^{(2)}}{\mu}\frac{\partial^2 {p_f^{(2)}}}{\partial{x_i}\partial{x_j}}+ \frac{k_{ij}^{(2,1)}}{\mu}\frac{\partial^2 {p_f^{(1)}}}{\partial{x_i}\partial{x_j}}\right)+\Gamma\left(p_f^{(1)}-p_f^{(2)}\right)
= \hphantom{XXXXXXXXXXXXX}\\
&\hphantom{XXXXXXXXXXXXX}\frac{1}{2}\sum^3_{i=1}\sum^3_{j=1}\alpha_{ij}^{(2)}\frac{\partial}{\partial{t}}\left(\frac{\partial{u_{i}}}{\partial{x_j}}+\frac{\partial{u_{j}}}{\partial{x_i}}\right)+\frac{1}{M^{(2)}}\frac{\partial{p_f^{(2)}}}{\partial{t}}+\frac{1}{M^{(1,2)}}\frac{\partial{p_f^{(1)}}}{\partial{t}}\,.
\end{aligned}
\end{equation} 
Three differential equations~(\ref{governing}) for stress distribution and two diffusion equations~(\ref{governing2a})--(\ref{governing2b}) for the fluid flow, determine five unknowns, $u_i$, $p_f^{(1)}$, and $p_f^{(2)}$. For $n$ pore-sets, the governing equations are rewritten as
\begin{equation}\label{governing3a}
\sum_{j=1}^3\left[\,\,\sum_{k=1}^3\sum_{\ell=1}^3\frac{1}{2}C_{ijk\ell}\left(\frac{\partial u_{k}}{\partial {x_j}\partial {x_\ell}}+\frac{\partial u_{\ell}}{\partial {x_j}\partial {x_k}}\right)-\sum_{p=1}^n\alpha^{(p)}_{ij}\frac{\partial{p_f^{(p)}}}{\partial{x_j}}\right]=0\,,
\end{equation}
\begin{equation}\label{governing3b}
\begin{aligned}
\sum_{i=1}^3&\sum_{j=1}^3\left( \frac{k_{ij}^{(p)}}{\mu}\frac{\partial^2 {p_f^{(p)}}}{\partial{x_i}\partial{x_j}}+\sum_{q=1}^n\frac{k_{ij}^{(p,q)}}{\mu}\frac{\partial^2 {p_f^{(q)}}}{\partial{x_i}\partial{x_j}}\right)-\sum_{q=1}^n\Gamma^{(p,q)}\left(p_f^{(p)}-p_f^{(q)}\right)
= \hphantom{XXXXXXXXXXXXX}\\
&\hphantom{XXXXXXXXXXXXX}\frac{1}{2}\sum^3_{i=1}\sum^3_{j=1}\alpha_{ij}^{(p)}\frac{\partial}{\partial{t}}\left(\frac{\partial{u_{i}}}{\partial{x_j}}+\frac{\partial{u_{j}}}{\partial{x_i}}\right)+\frac{1}{M^{(p)}}\frac{\partial{p_f^{(p)}}}{\partial{t}}+\sum_{q=1}^n\frac{1}{M^{(p,q)}}\frac{\partial{p_f^{(q)}}}{\partial{t}}\,.
\end{aligned}
\end{equation}  
There are $n+3$ equations that determine $n+3$ unknowns, $u_i$, $p_f^{(1)}, \dots\,, p_f^{(n)}$.
\section{Numerical examples}\label{sec:num}
Let us consider numerical examples to demonstrate the usage of consolidation equations. In our simulations, we go beyond the typical setting of single or dual porosity assumed in the past. Herein, we consider triple porosity, where a medium contains micropores, macropores, and fractures. Also, we allow different mechanical and diffusion properties of a pore set. The solutions of the partial differential equations were obtained using finite element methods incorporated inside the Matlab Reservoir Simulation Toolbox, MRST~\citep{LieMoyner21}. Within MRST, we generalized a dual-porosity module provided by~\citet{AshworthDoster19b} to fit our multi-porous extension. Although our examples have illustration purposes only, they are selected in such a way to mimic a probable geological scenario.

\subsection{Model conditions and parameters}
\begin{figure}[!htbp]
\includegraphics[scale=0.4]{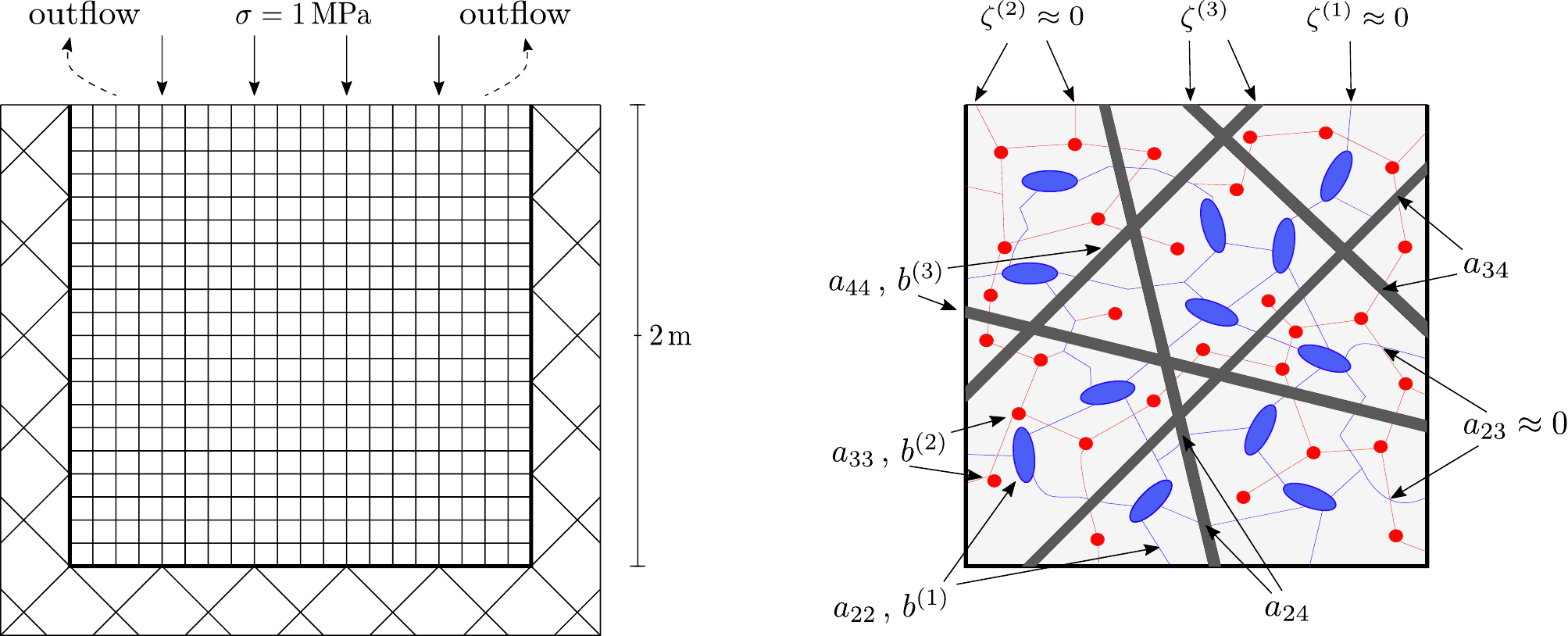}
\caption{\footnotesize{On the left: sketch of a geometrical setting, stress, and flow conditions. On the right: sketch of a triple-porosity scenario with a mapping of selected poroelastic coefficients. Connected macropores (set 1) are in blue, micropores (set 2) in red, and fractures (set 3) in black. Macro- and micro-porosity clusters are treated as isolated from each other, whereas the other sets are weakly connected. Ouflow coming from the background sets (macro- and micro-porosity) is neglected. }}
\label{fig:geom}
\end{figure}

We choose the following geometrical setting and stress-strain boundary conditions~(Figure~\ref{fig:geom}). The considered medium is $2$\,m$\times2$\,m with a regular $20\times20$ mesh. Changes in fluid pressures are induced by external compressional stress of $1$\,MPa imposed on the top side. Displacements are vertical only. The fluid is allowed to flow through the top boundary only. Medium is considered to be undrained at a time $t=0$. At the beginning of the consolidation, the material is intact and has the properties discussed in the paragraph below and schematically presented in Figure~\ref{fig:geom}.

This triple-porosity case can be typical for, e.g., reservoir rocks.
Although there is experimental evidence of multi-porosity behaviour, the exact measurement of poroelastic coefficients in geomaterials remains challenging. To the best of our knowledge, the storages and Biot-like parameters extracted in the laboratory are available for the simplified isotropic dual porosity case only~\citep[e.g.,][]{BerrymanWang95}. Therefore, similarly to other researchers, we must choose certain input values subjectively~\citep[e.g.,][]{ZhangEtAl18}. Our choices are based on the values proposed by~\citet{BerrymanPride02} for Weber sandstone. We assume three constituents, where the constituent (c) occupies a specific volume fraction $v^{(c)}$ and contains one pore set (p) with initial volume fraction $\phi^{(p)}<v^{(c)}$. Set 1 corresponds to macroporosity, set 2 describes microporosity, whereas set 3 refers to fractures. Material and fluid characteristics can be found in Table~\ref{tab:2}. The majority of parameters corresponding to sets 1 and 3 are taken from~\citet{BerrymanPride02}. The values of other coefficients---including all parameters of set 2---are chosen by us; they are bold in Table~\ref{tab:2}. 
\begin{table}[!htbp]
\begin{tabular}{ccccccccccc}
\multicolumn{3}{c}{Stiffnesses}   & \multicolumn{3}{c}{Poroel. coeff.} & \multicolumn{3}{c}{Fluid descr.}  & \multicolumn{2}{c}{\hphantom{XXX}Volume frac.\hphantom{XXX}}\\
\cmidrule{1-11}
$E^{(1)}$  & 21 &GPa & $\hphantom{xx}a_{22}$ & 0.0993 &GPa$^{-1}$   & \hphantom{xx}$k^{(1)}$ & {\bf{0}} &mD& \hphantom{xx}$v^{(1)}$  &  {\bf{0.5}}$\times$0.9905  \\          
$E^{(2)}$    &  {\bf{50}} &GPa  & $\hphantom{xx}a_{33}$  & {\bm{$x\,a_{22}$}} &GPa$^{-1}$ & \hphantom{xx}$k^{(2)}$ & {\bf{0}} &mD &\hphantom{xx} $v^{(2)}$  &    {\bf{0.49525}} \\
$E^{(3)}$    &  0.15& GPa    &   $\hphantom{xx}a_{44}$     & 0.145 &GPa$^{-1}$   & \hphantom{xx}$k^{(3)}$  & {\bm{$z$}} &mD & \hphantom{xx}$v^{(3)}$  &  0.0095   \\
$\nu^{(1)}$  &    0.15 &   &     $\hphantom{xx}b^{(1)}$    &  0.0253 &GPa$^{-1}$& \hphantom{xx}$k^{(1,2)}$  &{\bf{0}} &mD & \hphantom{xx}$\phi^{(1)}$ & 0.095  \\
$\nu^{(2)}$  &     {\bf{0.15}} &  &          \hphantom{xx}$b^{(2)}$    &  {\bm{$y\,b^{(1)}$}} &GPa$^{-1}$ & \hphantom{xx}$k^{(1,3)}$  & {\bm{$1/z$}} &mD &\hphantom{xx}$\phi^{(2)}$ & {\bf{0.05}} \\
$\nu^{(3)}$   &   {\bf{0.12}} &   &         \hphantom{xx} $b^{(3)}$             & 0.049 &GPa$^{-1}$   &  \hphantom{xx}$k^{(2,3)}$ & {\bm{$1/z$}} &mD &\hphantom{xx}$\phi^{(3)}$ & {\bf{0.009}} \\
$K_s$    &  37  &GPa      &                 \hphantom{xx} $a_{23}$       &     0 &TPa$^{-1}$   & \hphantom{xx}$\Gamma^{(1,2)}$  &{\bf{0}} &$\frac{1}{{\rm{GPa}}\times {\rm{s}}}$  &&  \\
$E_s$    &   {\bm{$62.9$}}  &GPa &   \hphantom{xx} $a_{24}$       &   {\bf{0.5}}$\times2.7$& TPa$^{-1}$ & \hphantom{xx}$\Gamma^{(1,3)}$  &{\bm{$1/z$}} &$\frac{1}{{\rm{GPa}}\times {\rm{s}}}$ &&  \\
$\nu_s$ &   {\bm{$0.15$}} &   &        \hphantom{xx}  $a_{34}$       &   {\bf{1.35}} &TPa$^{-1}$ &  \hphantom{xx}$\Gamma^{(2,3)}$  &{\bm{$1/z$}} &$\frac{1}{{\rm{GPa}}\times {\rm{s}}}$ &&  \\
& &&  &  && $\mu$  & {\bf{1}} &cP&&  \\
\cmidrule{1-11}
\end{tabular}
\caption{\small{Parameters for isotropic triple-porosity simulations. Specific values of unknowns $x$, $y$, $z$ correspond to different cases discussed. Bolded values are chosen subjectively by the authors. Other values were provided by~\citet{BerrymanPride02} based on laboratory measurements.}}
\label{tab:2}
\end{table}

Let us discuss the volume fractions and stiffnesses chosen. We assumed that the original volume occupied by unfractured porous background given by~\citet{BerrymanPride02} is equally divided between constituents with macro and micropores, $v^{(1)}=v^{(2)}$, where the macroporosity itself (set 1) occupies almost twice the space of the microporosity (set 2), $\phi^{(1)}\approx2\phi^{(2)}$. In general, Young's modulus diminishes with a higher concentration of pores~\citep[e.g.,][]{PabstGregorova14}. Due to a smaller void space in set 2, we choose $E_s>E^{(2)}>E^{(1)}$, where $E_s$ is Young's modulus of the solid phase, $E^{(2)}$ is the drained Young's modulus of the microporosity constituent, and $E^{(1)}$ denotes drained Young's modulus of the macroporosity constituent. The fractured constituent consists of a volume given by~\citet{BerrymanPride02}, where fractures occupy most of the phase that results in a very low Young's modulus, $E^{(3)}$. According to~\citet{LutzZimmerman21},  Poisson's ratio changes insignificantly when pores are spherical, and $\nu_s$ is close to $0.2$. On the other hand, if cracks are considered, Poisson's ratio should diminish. Therefore, we assumed that $\nu_s=\nu^{(1)}=\nu^{(2)}>\nu^{(3)}$.
To obtain the effective elasticity needed in the consolidation equations, we calculate effective Young's modulus $E$ and effective Poisson's ratio $\nu$ by employing a lower bound of Voigt's average, namely,
\begin{equation*}
E=\left(\phi^{(1)}\frac{1}{E^{(1)}}+\phi^{(2)}\frac{1}{E^{(2)}}+\phi^{(3)}\frac{1}{E^{(3)}}\right)^{-1}\approx11.5 {\rm{GPa}}\,,
\end{equation*}
\begin{equation*}
\nu=\left(\phi^{(1)}\frac{1}{\nu^{(1)}}+\phi^{(2)}\frac{1}{\nu^{(2)}}+\phi^{(3)}\frac{1}{\nu^{(3)}}\right)^{-1}=0.1282\,.
\end{equation*}

Now, let us discuss the values of coefficients describing fluids and poroelastic properties.
We chose the typical viscosity of water at $\mu=1$\,cP, and we selected permeabilities corresponding to the well-known situation of highly permeable fractures and background porosity having very low permeability. We assume that the time-dependent external diffusion of the fluid directly from the background is negligible that is reflected by $k^{(1)}=k^{(2)}=0$. Also, fluid does not exchange between micro and macro porosities, which is reflected by the intersection storage $a_{23}=0$, interflow permeability $k^{(1,2)}=0$, and leakage $\Gamma^{(1,2)}=0$. However, fluid exchange is allowed between micropores (set 1) and fractures (set 3) as well as between macropores (set 2) and fractures. Therefore, inter-pore sets permeabilities and leakages $k^{(1,3)}$, $k^{(2,3)}$, $\Gamma^{(1,3)}$ and $\Gamma^{(2,3)}>0$. To check the influence that the relative differences in  permeability and leakage between pore sets may inflict on  internal and external flux, we introduced multiplier $z$ equal to $10$, $10^3$, or $10^5$, which relate $k^{(p,q)}$ and $\Gamma^{(p,q)}$ as indicated in Table~\ref{tab:2}. Leakage coefficients governing the internal flow are obtained as follows.
\begin{equation*}
\Gamma^{(p,q)}=\frac{\delta \times k^{(p,q)}\,\left[{\rm{mD}}\right]}{\left(L\,\left[{\rm{m}}\right]\right)^2\times\mu\,\left[{\rm{cP}}\right] }=\frac{\pi^2k^{(p,q)}\times9.869233\times10^{-16}\,[{\rm{m}}^2]}{\mu\times10^{-5}\,[{\rm{Pa}}\times {\rm{s}}\times {\rm{m}}^2]}\approx k^{(p,q)}\,\left[\frac{1}{{\rm{GPa}}\times {\rm{s}}}\right]\,,
\end{equation*}
where $\delta$ is the shape factor~\citep{WarrenRoot63} assumed to be equal to $\pi^2$, and $L=0.1\rm{m}$ is the fracture spacing viewed as the parameter that corresponds to the fracture concentration. 
To check the influence of the relative difference between pore sets in poroelastic coefficients, $a_{ij}$, i.e., storages, and $b^{(p)}$, i.e., poroelastic expansion coefficients, we introduced multipliers $x$ and $y$ that are respectively equal to $0.8$ and $0.5$, $0.5$ and $0.5$, or $0.5$ and $0.8$. Based on the work of~\citet{SelvaduraiSuvorov20}, the Biot coefficient is usually lower for a lower concentration of pores that have similar shapes. Combining~(\ref{b_alt1}) and~(\ref{biotwillis}), we can express the isotropic expansion coefficient as,
\begin{equation*}
b^{(p)}=\frac{1}{3K}\alpha^{(p)}.
\end{equation*}
Hence, $b^{(p)}$ is strictly related to the Biot-like coefficient; its value is expected to be lower for lower pore concentrations. This is why the chosen values of $y$ are below 1 to relate poroelastic expansion coefficients between macro and micropores (sets 1 and 2, respectively). In the literature, the values of fluid storage are still weakly explored, but we suspect achieving lower storage for lower pore concentrations of similar shape, which is why $x$ is also below 1. Since the shape of the inhomogeneity can greatly affect the poroelastic coefficient~\citep[e.g.,][]{SelvaduraiSuvorov20}, the interplay of $x$ and $y$ can mimic different pore configurations between background constituents, i.e., macro and micro pore sets.

\subsection{Results}\label{sec:num2}
\begin{figure}[!htbp]
\centering
\begin{subfigure}{.3\textwidth}
\includegraphics[scale=0.4]{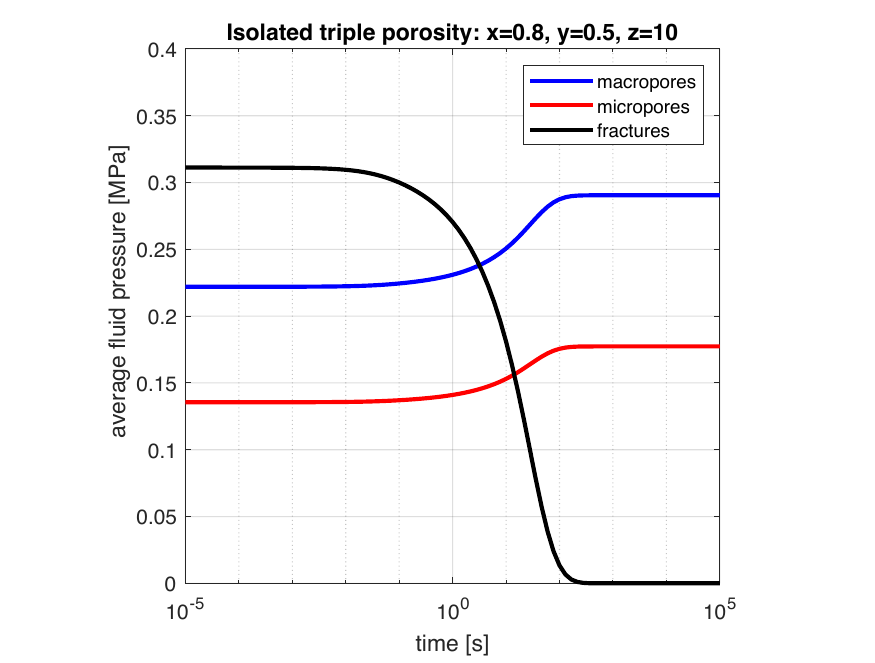}
\caption{\footnotesize{}}
\label{fig:isoa}
\end{subfigure}
\begin{subfigure}{.3\textwidth}
\includegraphics[scale=0.4]{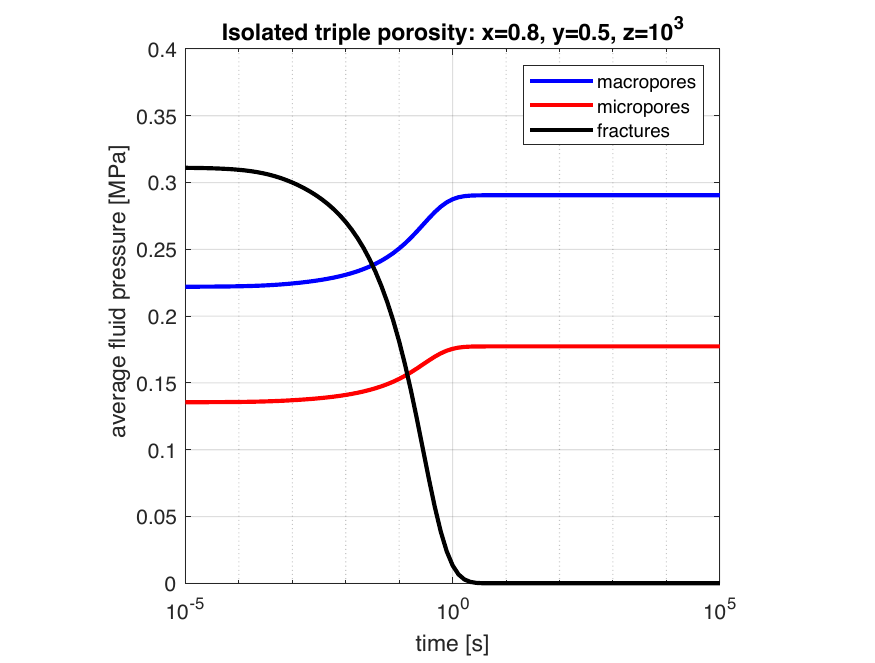}
\caption{\footnotesize{}}
\label{fig:isob}
\end{subfigure}
\begin{subfigure}{.3\textwidth}
\includegraphics[scale=0.4]{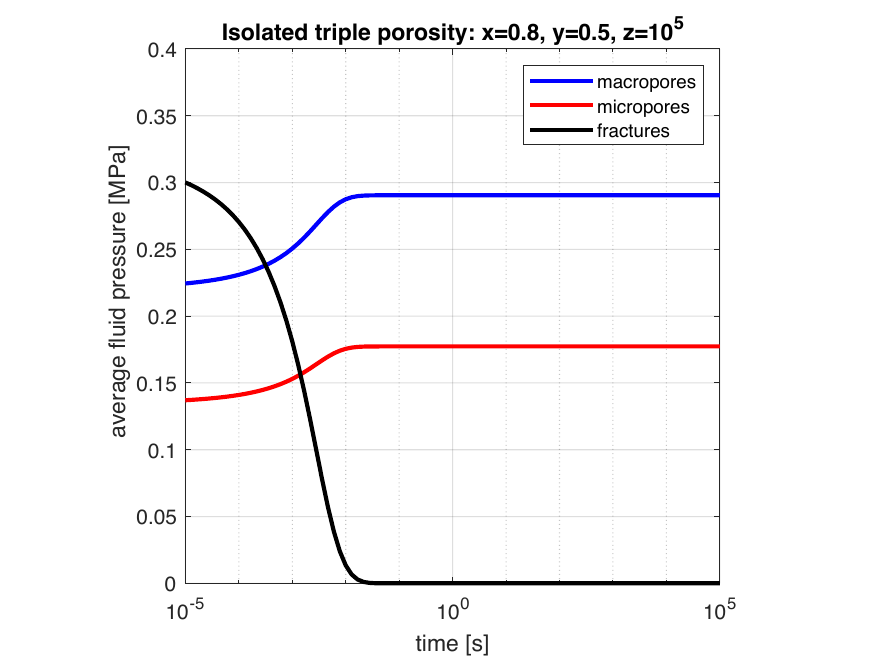}
\caption{\footnotesize{}}
\label{fig:isoc}
\end{subfigure}
\caption{\footnotesize{Semi-log plots of the average fluid pressure changes with time. Initially undrained triple-porosity medium with isolated sets. Microporosity and macroporosity remain undrained, whereas the fractured constituent is drained with time. } }
\label{fig:iso}
\end{figure}

In this section, we discuss the results of the triple-porosity simulations. Specifically, we focus on the time-dependent fluid pressure changes caused by the fluid outflow and gradual drainage of the medium. Before we move to the results of the model configurations discussed previously, first, let us invoke an end member case of isolated pore sets. In other words, assume that the drainage is achieved for the fractured constituent only, whereas the background porosity clusters (sets 1 and 2) remain undrained. To obtain such scenario, we set $a_{23}=a_{24}=a_{34}=0$, $k^{(1,2)}=k^{(1,3)}=k^{(2,3)}=0$, and $\Gamma^{(1,2)}=\Gamma^{(1,3)}=\Gamma^{(2,3)}=0$.
The rest of the parameters from Table~\ref{tab:2} remain unchanged. The results are provided for $x=0.8$, $y=0.5$, and various values of $z$; they are illustrated in Figure~\ref{fig:iso}.

We notice that the solid deformation and the drainage of fractures lead to an increase in fluid pressures of the background constituents and a decrease in fracture fluid pressure. The above phenomenon can be explained as follows. In the absence of fluid volume change within the porous background, $\zeta^{(1)}=\zeta^{(2)}=0$, solid compression leads to the increase of a volumetric strain, $e$, that must be compensated by the increase of $p_f^{(1)}$ and $p_f^{(2)}$; see e.g., equation~(\ref{mixedzeta}). Further, the increased aforementioned pressures equilibrize the decrease of $p_f^{(3)}$ caused by the fluid outflow from the fractures; see e.g., equation~(\ref{version3}).
Due to the isolation of the background sets, their pressures must increase till the drainage process is finished. Naturally, the complete drainage appears earlier for larger $k^{(3)}$. This is why $p_f^{(3)}$ approaches zero most rapidly in Fig~\ref{fig:isoc}, where $k^{(3)}=10^5$\,mD, and most slowly in Fig~\ref{fig:isoa}, where $k^{(3)}=10$\,mD.

\begin{figure}[!htbp]
\centering
\begin{subfigure}{.3\textwidth}
\includegraphics[scale=0.4]{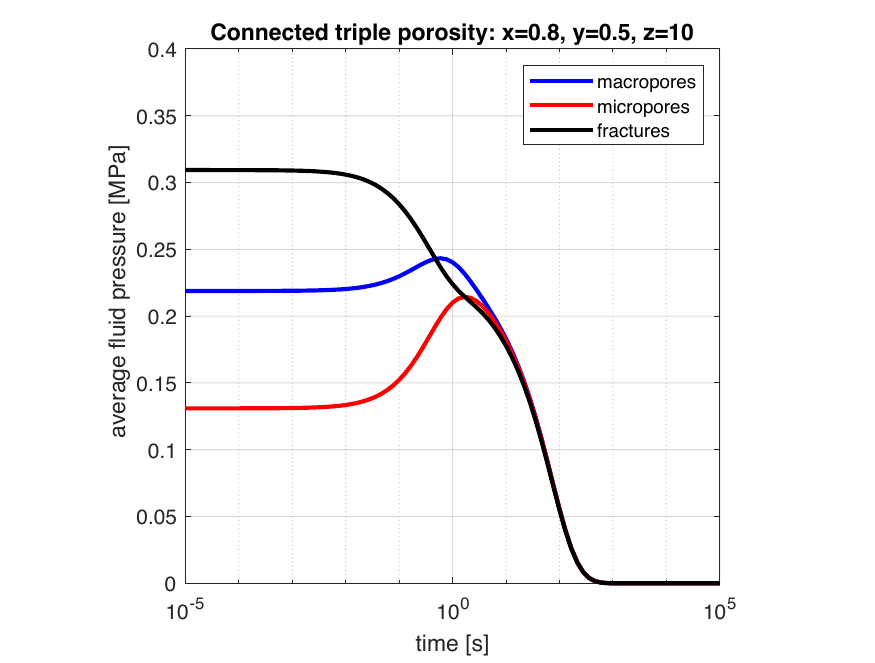}
\caption{\footnotesize{}}
\label{fig:cona}
\end{subfigure}
\begin{subfigure}{.3\textwidth}
\includegraphics[scale=0.4]{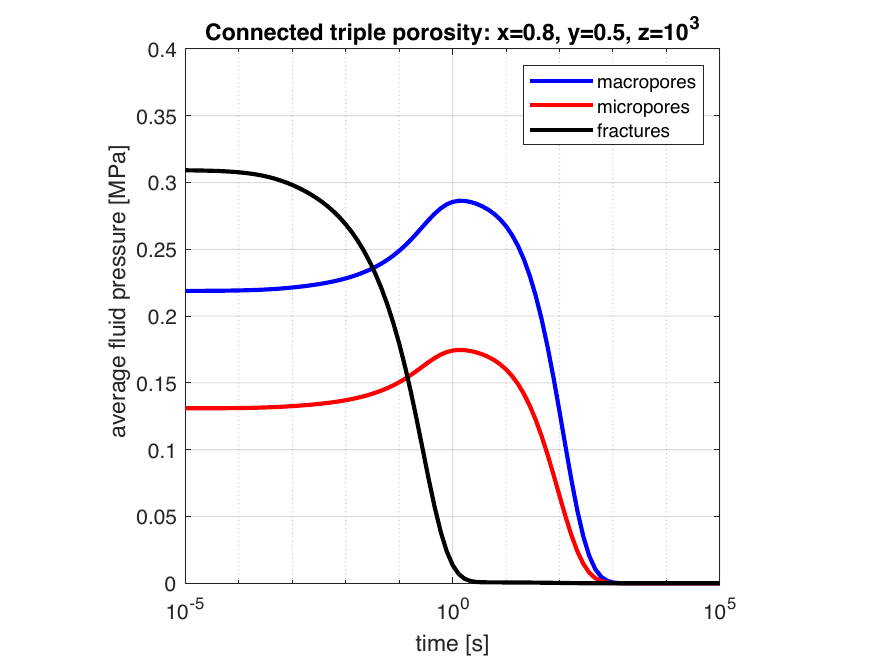}
\caption{\footnotesize{}}
\label{fig:conb}
\end{subfigure}
\begin{subfigure}{.3\textwidth}
\includegraphics[scale=0.4]{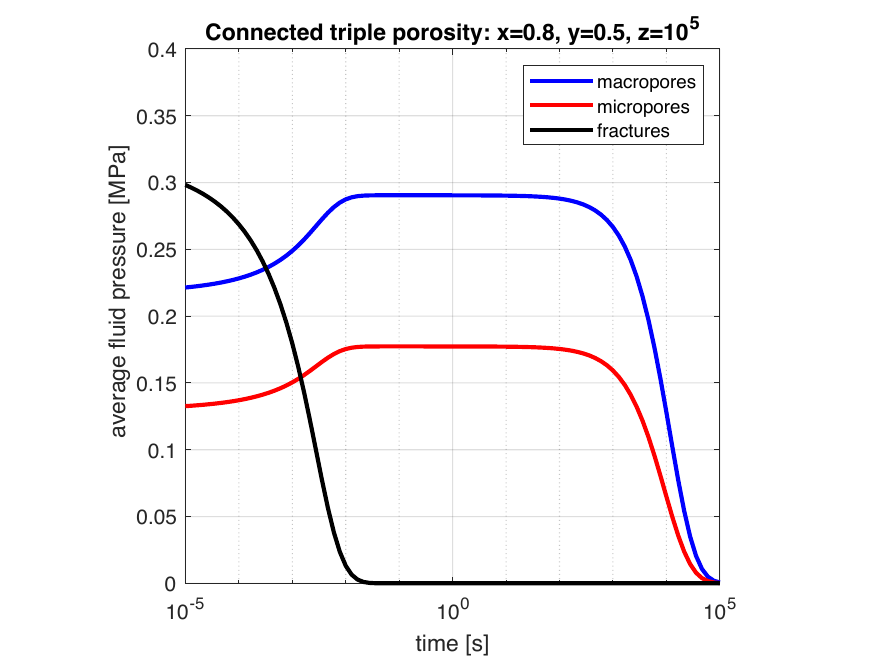}
\caption{\footnotesize{}}
\label{fig:conc}
\end{subfigure}
\begin{subfigure}{.3\textwidth}
\includegraphics[scale=0.4]{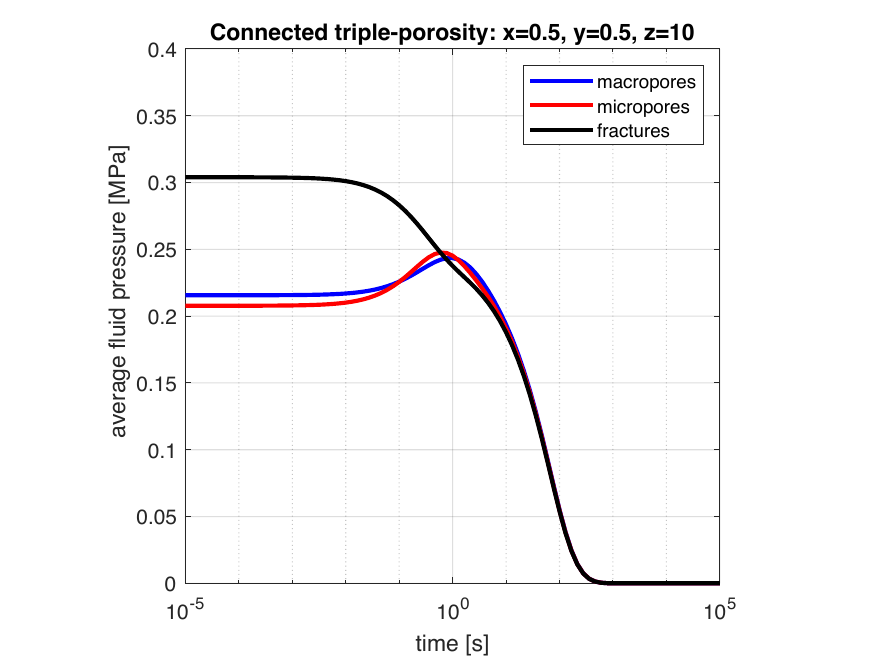}
\caption{\footnotesize{}}
\label{fig:cond}
\end{subfigure}
\begin{subfigure}{.3\textwidth}
\includegraphics[scale=0.4]{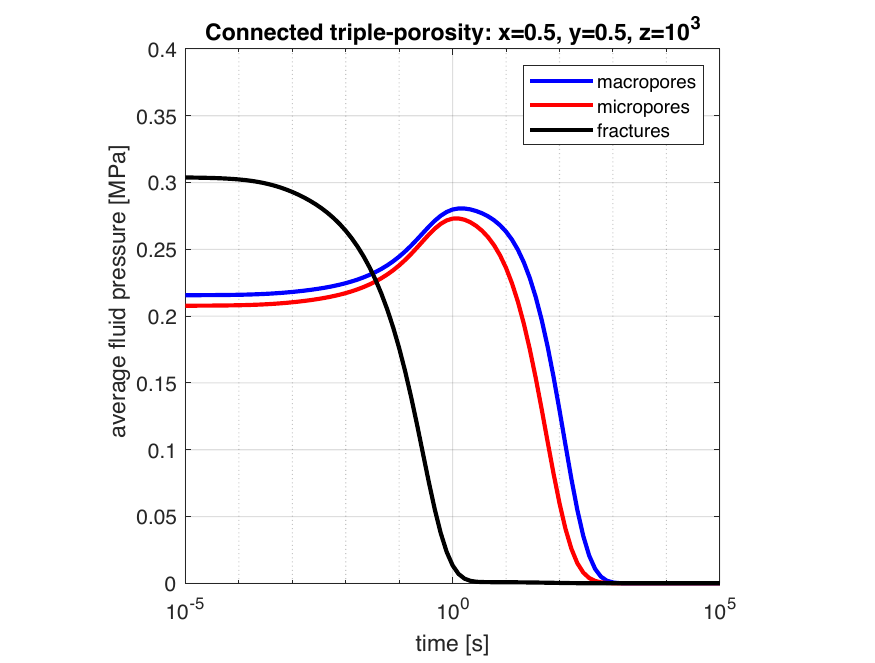}
\caption{\footnotesize{}}
\label{fig:cone}
\end{subfigure}
\begin{subfigure}{.3\textwidth}
\includegraphics[scale=0.4]{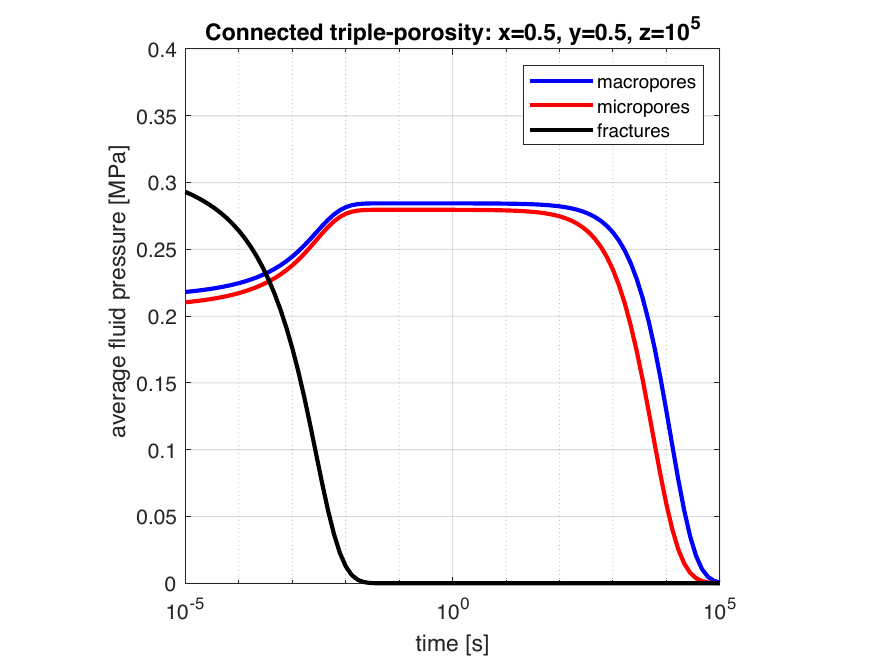}
\caption{\footnotesize{}}
\label{fig:conf}
\end{subfigure}
\begin{subfigure}{.3\textwidth}
\includegraphics[scale=0.4]{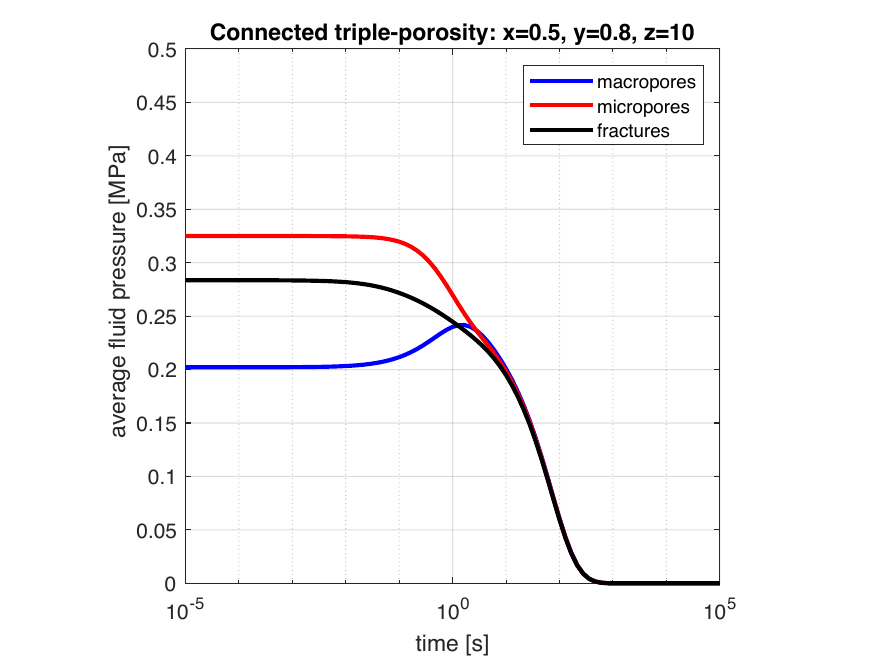}
\caption{\footnotesize{}}
\label{fig:cong}
\end{subfigure}
\begin{subfigure}{.3\textwidth}
\includegraphics[scale=0.4]{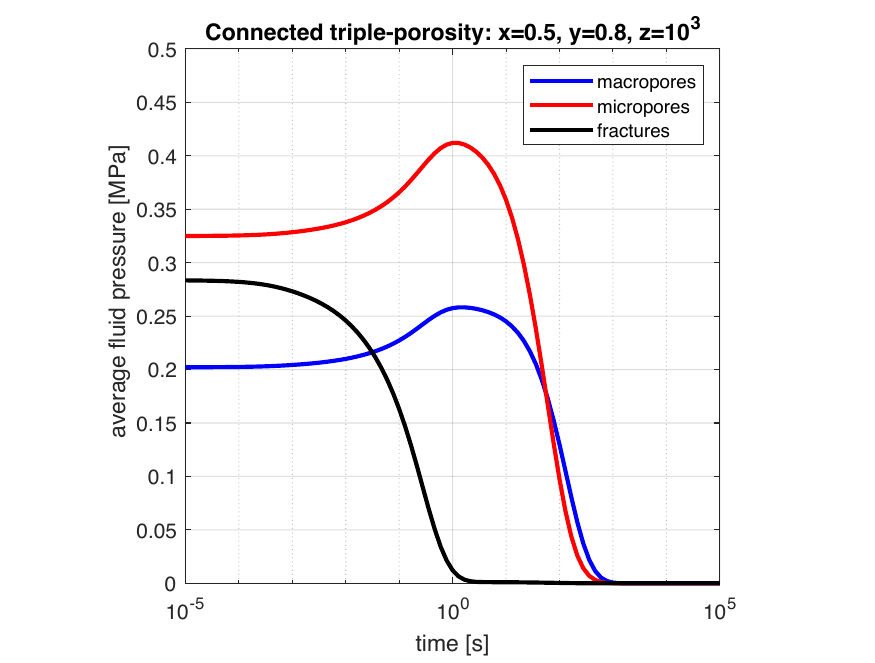}
\caption{\footnotesize{}}
\label{fig:conh}
\end{subfigure}
\begin{subfigure}{.3\textwidth}
\includegraphics[scale=0.4]{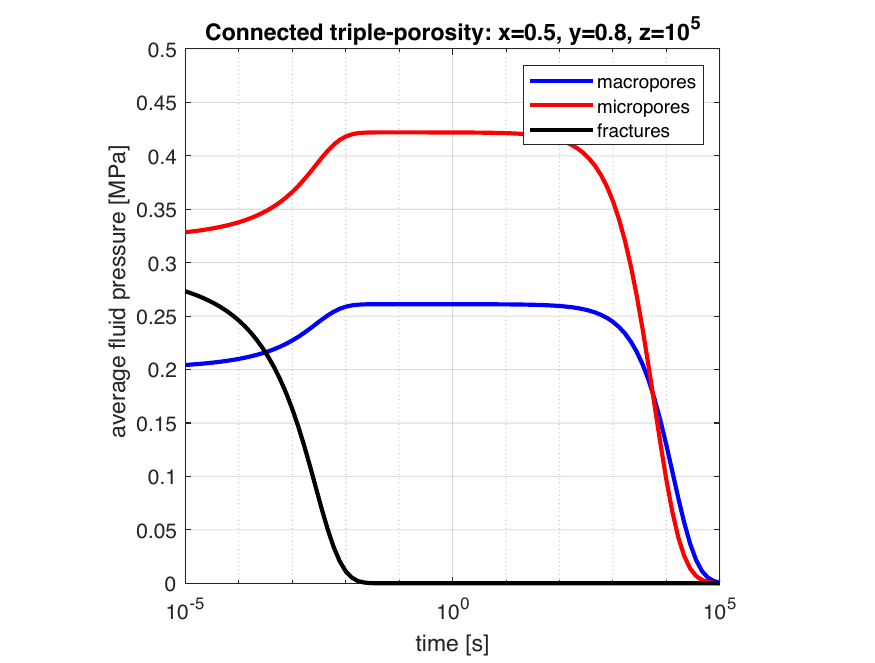}
\caption{\footnotesize{}}
\label{fig:coni}
\end{subfigure}
\caption{\footnotesize{Semi-log plots of the average fluid pressure changes with time. Initially undrained triple-porosity medium with weakly-connected sets. Each set becomes drained with time.
} }
\label{fig:con}
\end{figure}

Now, let us consider examples with the input values given in Table~\ref{tab:2}. In other words, we simulate the scenario of weakly-connected pore sets. Results for all combinations of $x$, $y$, and $z$ multipliers are provided in Figure~\ref{fig:con}. Particularly, Figures~\ref{fig:cona}--\ref{fig:conc}, have the same multipliers as in the previous case of isolated sets so that the isolated and connected scenarios can be compared directly.

Let us discuss the results. In Figures~\ref{fig:cona}--\ref{fig:conc}, background pressures increase gradually and, upon a certain time, they start to decrease abruptly. The decrease appears earlier for larger interflow permeabilities (stronger leakage). If the interflow is small, as illustrated in Figure~\ref{fig:conc}, pressures converge with the isolated sets scenario from Figure~\ref{fig:isoc} up to $t\approx 10$\,s, after which macro and micropore fluid pressure abruptly decrease to zero. By contrast, we notice that a relatively small contrast in fracture and interflow permeabilities leads to pressure equilibration and single-porosity response over a short period of time, as shown in Figure~\ref{fig:cona}. An interesting bump of background pressures can be noticed in Figure~\ref{fig:conb} at $t\approx 1$\,s. This can be explained as follows. Initially, some fluid volume contained in fractures drains directly from them, and background sets behave as isolated since their pressures are lower than that in fractures, which impedes internal leakage (the background is already fully saturated). Subsequently, once pressures in the background become higher, the fluid starts to diffuse towards the fractures so that the pressures equilibrize throughout the medium. However, such an equilibration process does not happen instantaneously due to the much lower interflow permeability that causes the bump. In the case of low contrast in permeabilities, the bump is hardly visible. Figures~\ref{fig:cond}--\ref{fig:conf}, display the case in which multipliers $x=y=0.5$ and therefore represent a higher contrast in poroelastic coefficients between micro and macro-pores sets. This situation exhibits similar pressure evolution to that shown in Figures~\ref{fig:cona}--\ref{fig:conc}, but with very small difference between background (macro and micro pore sets) pressures. This small contrast in background pressures is representative of an almost dual-porosity response. This phenomenon of nearly dual-porosity behavior i.e., micro and macro pores acting as a single set, may appear counter-intuitive considering that both storage and Biot-like coefficients are greatly diminished for the microporosity ($x=y=0.5$) compared to the macroporosity. However, it can be explained as follows. The absence of the external outflow from the background, $\zeta^{(1)}=\zeta^{(2)}=0$, can be rewritten using~(\ref{multiporaniso}) along with $a_{33}=xa_{22}$, and $b^{(2)}=yb^{(1)}$ as
\begin{align*}
0=b^{(1)}\sigma+a_{22}p^{(1)}+a_{23}p^{(2)}+a_{24}p^{(3)}\,,\\
0=yb^{(1)}\sigma+a_{23}p^{(1)}+xa_{22}p^{(2)}+a_{34}p^{(3)}\,.
\end{align*}
Knowing that intersecting storages are zero or very low, then we get
\begin{equation}\label{relativ}
b^{(1)}\sigma+a_{22}p^{(1)}\approx
\frac{y}{x}b^{(1)}\sigma+a_{22}p^{(2)}\,
\end{equation} 
therefore, if $x=y$, then  $p^{(1)}\approx p^{(2)}$.  
In consequence, the ratio between $x$ and $y$---not just their absolute values---influences substantially the relationship between micro and macro-porosity pressure evolution.
The importance of poroelastic coefficients in the consolidation process is again confirmed in Figures~\ref{fig:cong}--\ref{fig:coni}. These figures represent a case of relatively low storage and high Biot-like coefficients that cause high fluid pressure in microporosity.

Our simulations show the consolidation process in multi-porous media and provide certain insights on time-dependent changes in fluid pressures. We demonstrated that the pressure evolution and pore sets interplay is controlled by the relative differences in permeability and poroelastic coefficients between pore sets. By considering different combinations of pore sets permeabilities and poroelastic coefficients, we showed specific scenarios where the triple-porosity case converges to single or dual-porosity responses. 
In particular, a triple porosity medium will rapidly converge to a single porosity medium if the contrast between the permeability of the most permeable pore set (fractures in this example) and the inter-pore sets permeability is small (less than two orders of magnitude difference in this example) (Figures~\ref{fig:cona}, \ref{fig:cond}, \ref{fig:cong}). We also identified characteristic features, such as pressure ``bumps'' (Figures~\ref{fig:conb}, \ref{fig:cone} ,\ref{fig:conh}) appearing in the consolidation process when the contrast between fracture and inter-pore sets permeability is sufficient to cause distinctive responses between pore sets but not large enough to cause pressure equilibration in the background for a significant amount of time before draining (Figures~\ref{fig:conc}, \ref{fig:conf}, \ref{fig:coni}).
Our examples should not be treated as conclusive but rather encourage future investigations on other multi-porous scenarios. 
\section{Discussion}
Our multi-porous extension provides a rigorous description of an instantaneous deformation along with time-dependent fluid flow and consolidation of a solid medium containing complex porous structures. The effects of pore sets and their anisotropic responses are not neglected.
To obtain such an accurate description that accounts for mesoscopic inhomogeneities (pore-sets), many coefficients are needed (see e.g., Table~\ref{tab:2}). These may be difficult to obtain in both laboratory and field measurements. Nevertheless, recent studies~\citep{BrantutAben21} revealed that an accurate determination of local variations in fluid pressure, at the scale of the laboratory sample, is possible. Given rapid developments in experimental geophysics, one should optimistically look to the future applicability of the multi-porous extension. Further, alternative tests to the ones from Sections~\ref{sec4}--\ref{sec5} are possible. First, we can consider uniform-expansion thought experiments~\citep{Berryman02}. Second, the poroelastic coefficients can be obtained from micromechanical derivations, as presented in our parallel paper~\citep{AdamusEtAl23b}. In the planned future developments, we will perform the experiments proposed in Sections~\ref{sec4}--\ref{sec5}. 

Our analysis of time-dependent deformation for multi-porous anisotropic poroelasticity reveals potentially significant mechanical consequences. 
For both isolated~(Figure~\ref{fig:iso}) and weakly-connected~(Figure~\ref{fig:con}) pore sets (1 and 2), the pore fluid pressure is predicted to rise as the pore fluid drains through the fractures (set 3). In the case of weakly-connected pore sets 1 and 2 shown in Figure~\ref{fig:con}, there are distinct transient peaks in pore fluid pressure of the order of several seconds duration. 
This means that in a draining poroelastic material with multiple pore sets such as rock loaded in a fault zone, as the fluid drains through the high permeability fractures and the pore fluid pressure drops therein, the pore fluid pressure in weakly-connected pore sets in the rock matrix increases for short periods (seconds), thereby reducing the local effective stress $\sigma_{eff}=\sigma+p_f^{(p)}$.
The pore fluid pressure decrease in the fractures causes a local increase in effective stress. 
These localized changes in effective stress will push the matrix containing the weakly connected pore sets towards mechanical failure, and the fractures away from mechanical failure~\citep[e.g.,][]{Lockner95}. 
Moreover, it is clear from the cases modeled in Figure~\ref{fig:con} that the magnitude and duration of the pore fluid pressure transient in the weakly-connected pore sets (1 and 2)---and therefore the period in which the effective stress is reduced---critically depends on their relative poroelastic properties ($a_{33}$, $b^{(2)}$).
Further analysis of this phenomenon is beyond the scope of the present work but will be addressed in a follow-up paper.

Two main alternatives to the presented multi-porous theory extension can be used to model the mechanical behaviour of complex porous structures; homogenization and discretization~\citep{AshworthDoster19}. The medium can be homogenized and treated as poro-elastically effective~\citep{Berryman06} or discretized, where local subgrid flows are determined individually~\citep{KarimiFardEtAl06}. However, homogenization does not allow us to distinguish different flow characteristics. Using elastic layers as an analogy, the homogenisation of thin constituents makes sense from the long-wave seismic perspective, but the intrinsic properties of layers are lost. The discretisation is difficult to introduce at a field scale, where fractures are abundant~\citep{AshworthDoster19}, and the integration of the subgrids to a continuum model presented herein can cause issues~\citep{Berkowitz02}.
If pore sets are spatially distributed (not nested, not fractured), one can try to discretise the medium to obtain a local, single---instead of multiple---porosity case (Figure~\ref{fig:one}). Nevertheless, if extended coefficients can be either measured or estimated, we recommend using the multi-porous extension as the most accurate macro-meso-mechanical description.

It seems that there is no best choice of the number of pore sets, $n$. We believe that in reality, at a microscale and in a short-time period, each pore within the same pore set can be characterized by slightly or even significantly different poroelastic properties. The connections between pores in a pore set are not ideal and hence, most probably, lead to pressure, permeability, and fluid increment pore-scale variations. Thus, one can choose between macroscopic single-porosity description ($n=1$), more accurate---but still somewhat idealized---mesoscopic dual-porosity description ($n=2$),  or even more accurate mesoscopic triple-porosity description ($n=3$), and so on ($n\geq4$), till fully microscopic description is reached ($n=$total number of pores)~\citep{AdamusEtAl23b}. Naturally, the choice of the model is a trade-off between scale refinement and the accuracy of the estimation---if at all possible---of the parameters.
\section{Summary}
We have proposed the multi-porous extension of anisotropic poroelasticity---quasi-static theory originally presented by~\citet{Biot41}. In the extended theory, pores or cracks form multiple sets (porosity clusters). Such sets are either weakly connected or isolated from each other, leading to non-uniform fluid content changes and fluid pressure changes within a bulk medium. Each set may induce anisotropy, meaning that pores in a set are not necessarily randomly oriented. Also, the solid matrix---in which the sets are embedded---is allowed to be anisotropic. 

In Section~\ref{sec2}, we have indicated practical scenarios pertinent to our extension: hierarchical porosity, complex porosity, and spatially-distributed pore sets. Also, we referred to the cases where a simpler theoretical extension of isotropic dual porosity was already utilized with satisfactory results.
In Section~\ref{sec3}, we formulated the strain-stress relations that employed novel poroelastic coefficients $a_{ij}$ and $b_{ij}^{(p)}$.
In Section~\ref{sec4}, we determined the physical meaning of these coefficients. $a_{ij}$ can be viewed as storages under constant stress, whereas $b_{ij}^{(p)}$ are the poroelastic expansions or products of storages and Skempton-like coefficients. Several types of tests to obtain $a_{ij}$ and $b_{ij}^{(p)}$ were proposed. 
In Section~\ref{sec5}, we discussed alternative formulations of strain-stress relations. Alternative coefficients $1/M^{(p)}$ and $\alpha_{ij}^{(p)}$ were proposed and determined. $1/M^{(p)}$ can be viewed as storages under no frame deformation, whereas $\alpha_{ij}^{(p)}$ are the Biot-like coefficients. Relevant types of laboratory experiments were indicated.
Finally, in Section~\ref{sec6}, we introduced time dependency to obtain the governing equations of three-dimensional consolidation~(\ref{governing3a})--(\ref{governing3b}). To do so, we proposed the extended Darcy's law and fluid continuity equations. These equations were combined with mixed strain-stress formulations to obtain novel diffusion equations. The remaining governing equations were derived from the stress equilibrium conditions. 
In Section~\ref{sec:num}, the usage of consolidation equations was demonstrated on novel simulations of the triple-porosity case. These numerical examples demonstrated that the pore sets pressure evolution and interplay is controlled by the relative differences in permeability and poroelastic coefficients between pore sets. The simulations of the time-dependent drained behaviour of a multi-porous anisotropic poroelastic material show that positive pore pressure transients are generated in the weakly connected pore sets, and these could potentially be of sufficient magnitude and duration to push the material towards brittle failure. 

\section*{Acknowledgements}
This research was supported financially by the NERC grant: ``Quantifying the Anisotropy of Poroelasticity in Stressed Rock'', NE/N007826/1 and NE/T00780X/1.
\section*{Data availability}
The data that support the findings of this study are available from the corresponding author upon reasonable request.
\section*{Conflict of interest}
The authors have no conflict of interest to declare.
\bibliographystyle{apa}
\bibliography{LibraryUpperCase}
\appendix
\addcontentsline{toc}{section}{Appendices}
\section{List of symbols}\label{ap2}
\begin{table}[!htbp]
\hspace*{-6.2cm}
\scalebox{0.8}{
\begin{tabular}
{l c l}
\multicolumn{3}{l}{\bf{Bulk medium perspective}}\\ 
[2ex]
\multicolumn{3}{l}{Constant properties}\\
[2ex]
$\Delta_{ijk\ell}$&:=& effect of fluid in a medium       \\
$\delta$&:=& shape factor inside leakage coefficient                      \\
$\mu$&:=& fluid viscosity                      \\
$\nu$&:=& Poisson's ratio of a drained medium                      \\
$\nu_s$&:=& Poisson's ratio of a solid phase                      \\
$B_{ij}$&:=& Skempton coeff. of a medium              \\
$C_{ijk\ell}$&:=& drained elasticity tensor                \\
 $C^u_{ijk\ell}$&:=& undrained elasticity tensor         \\
$E$&:=& Young's modulus of a drained medium                      \\
$E_s$&:=& Young's modulus of a solid phase                      \\
$K$&:=& bulk modulus of a drained medium                      \\
$K_s$&:=& bulk modulus of a solid phase                      \\
$L$&:=& fracture spacing                      \\
$S_{ijk\ell}$&:=& drained compliance tensor                  \\
$S^u_{ijk\ell}$&:=& undrained compliance tensor      \\
$S$&:=& fluid storage in a medium                 \\
$V$&:=& volume of a medium        \\
[2ex]
\multicolumn{3}{l}{Variables} \\
[2ex]
$\zeta$&:=& fluid content change in a single-porosity medium\\
$\zeta_{tot}$&:=& total fluid content change in a medium\\
 $\varepsilon_{ij}$&:=& strain tensor \\
  $\sigma_{ij}$&:=& stress tensor\\
 $p_c$&:=& change of confining pressure       \\
$p_f$&:=& change of fluid pressure         \\
$U_i$&:=& displacement of a fluid contained in a medium\\
 $u_{i}$&:=& displacement of a solid skeleton                \\
[2ex]
\multicolumn{3}{l}{\bf{Mesoscopic perspective}}\\ 
[2ex]
\multicolumn{3}{l}{Constant properties} \\
[2ex]
$\alpha_{ij}^{(p)}$&:=& Biot-like coeff. of a $p$-th set          \\
 $\Gamma^{(p,q)}$&:=& interset leakage between $p$-th and $q$-th sets  \\
$\Delta^{(p)}_{ijk\ell}$&:=& effect of fluid of a $p$-th set                    \\
 $\nu^{(c)}$&:=& Poisson's ratio of a drained constituent containing a set       \\
$\phi^{(p)}$&:=& volume fraction of a $p$-th set \\
$a_{p+1,p+1}$&:=& constant-stress storage of a $p$-th set     \\
 $a_{p+1,q+1}$&:=& constant-stress storage at sets' intersection                   \\
 $b^{(p)}_{ij}$&:=& expansion coeff. of a $p$-th set                  \\
$B^{(p)}_{ij}$&:=& Skempton-like coeff. of a $p$-th set      \\
 $E^{(c)}$&:=& Young's modulus of a drained constituent containing a set       \\
$k^{(p)}_{ij}$&:=& permeability coeff. of a $p$-th set     \\
 $M^{(p)}$&:=& no-frame-deformation storage of a $p$-th set           \\
 $M^{(p,q)}$&:=& no-frame-deformation storage at sets' intersection      \\
 $N^{(p)}$&:=& relative storage of a $p$-th set                    \\
 $N^{(p,q)}$&:=& relative storage at sets' intersection         \\
 $S^{(p)}$&:=& constant-stress storage of a $p$-th set                    \\
 $S^{(p,q)}$&:=& constant-stress storage at sets' intersection        \\
 $V^{(p)}$&:=& volume of a set        \\
 $v^{(c)}$&:=& volume fraction of a constituent containing a set       \\
[2ex]
\multicolumn{3}{l}{Variables} \\
[2ex]
 $\zeta^{(p)}$&:=& external fluid increment to $p$-th set \\
$\zeta^{(p,q)}$&:=& interset fluid increment between $p$-th and $q$-th sets \\
$p_f^{(p)}$&:=& change of fluid pressure in $p$-th set                                \\
 $q^{(p)}$&:=& fluid flux to a $p$-th set         \\
$U_i^{(p)}$&:=& displacement of a fluid contained in a $p$-th set          \\
\end{tabular}
}
\end{table}
\section{Stress-strain relations for dual porosity}\label{ap1}
Herein, we rewrite equations~(\ref{version3})--(\ref{mixedzeta}) to obtain stress-strain relations for dual porosity. 
We get
\begin{align*}
\sigma_{ij}&=\sum^3_{k=1}\sum^3_{\ell=1}\left(C^u_{ijk\ell}-2N^{(1,2)}\alpha_{ij}^{(1)}\alpha_{k\ell}^{(2)}\right)e_{k\ell}+\left(N^{(1,2)}\alpha_{ij}^{(2)}-N^{(1)}\alpha_{ij}^{(1)}\right)\zeta^{(1)}+\left(N^{(1,2)}\alpha_{ij}^{(1)}-N^{(2)}\alpha_{ij}^{(2)}\right)\zeta^{(2)}\,,\\ 
p_f^{(1)}&=\sum^3_{k=1}\sum^3_{\ell=1}\left(N^{(1,2)}\alpha_{k\ell}^{(2)}-N^{(1)}\alpha_{k\ell}^{(1)}\right)e_{k\ell}+N^{(1)}\zeta^{(1)}-N^{(1,2)}\zeta^{(2)}\,,\\
p_f^{(2)}&=\sum^3_{k=1}\sum^3_{\ell=1}\left(N^{(1,2)}\alpha_{k\ell}^{(1)}-N^{(2)}\alpha_{k\ell}^{(2)}\right)e_{k\ell}+N^{(2)}\zeta^{(2)}-N^{(1,2)}\zeta^{(1)}\,,
\end{align*}
where 
\begin{equation*}
N^{(1)}:=\frac{M^{(1)}M^{(1,2)}M^{(1,2)}}{M^{(1,2)}M^{(1,2)}-M^{(1)}M^{(2)}}\,,
\end{equation*}
\begin{equation*}
N^{(2)}:=\frac{M^{(2)}M^{(1,2)}M^{(1,2)}}{M^{(1,2)}M^{(1,2)}-M^{(1)}M^{(2)}}\,,
\end{equation*}
\begin{equation*}
N^{(1,2)}:=\frac{M^{(1)}M^{(2)}M^{(1,2)}}{M^{(1,2)}M^{(1,2)}-M^{(1)}M^{(2)}}\,
\end{equation*}
are the coefficients described by the combinations of storages.
If the two sets are isolated from each other, then $1/M^{(1,2)}=0$. In such a case $N^{(1)}=M^{(1)}$, $N^{(2)}=M^{(2)}$, $N^{(1,2)}=0$ and the above stress-strain relations reduce to (\ref{version4})--(\ref{version4b}), as expected.
\end{document}